\documentclass[a4paper,hidelinks, 12pt]{article}
\usepackage[utf8]{inputenc}
\usepackage{a4wide}
\usepackage{graphics}
\usepackage{amsmath}
\usepackage{amsfonts}
\usepackage{amssymb}
\usepackage{amsthm}

\usepackage{subfigure}
\usepackage{cite}
\usepackage{xcolor}
\usepackage{soul}
\usepackage{cancel}
\usepackage[toc,page]{appendix}
\numberwithin{equation}{section}
\usepackage{mathtools}
\usepackage{hyperref}
\usepackage{enumitem}
\usepackage{multirow,multicol}
\usepackage{threeparttable}
\usepackage{etoolbox}

\allowdisplaybreaks

\let\sillymacro\theequation 
\patchcmd\sillymacro{equation}{parentequation}{}{}

\newcommand{\hc}{{\rm h.c.}}
\newcommand{\diag}{{\rm diag}}

\newtheorem{theorem}{Theorem}

\begin{document}

\begin{titlepage}
\begin{center}

{\large \bf {Complex \boldmath$S_3$-symmetric 3HDM}}

\vskip 1cm

A. Kun\v cinas, $^{a,}$\footnote{E-mail: Anton.Kuncinas@tecnico.ulisboa.pt}
O. M. Ogreid,$^{b,}$\footnote{E-mail: omo@hvl.no}
P. Osland$^{c,}$\footnote{E-mail: Per.Osland@uib.no} and 
M. N. Rebelo$^{a,}$\footnote{E-mail: rebelo@tecnico.ulisboa.pt}

\vspace{1.0cm}

$^{a}$Centro de F\'isica Te\'orica de Part\'iculas, CFTP, Departamento de F\'\i sica,\\ Instituto Superior T\'ecnico, Universidade de Lisboa,\\
Avenida Rovisco Pais nr. 1, 1049-001 Lisboa, Portugal,\\
$^{b}$Western Norway University of Applied Sciences,\\ Postboks 7030, N-5020 Bergen, 
Norway, \\
$^{c}$Department of Physics and Technology, University of Bergen, \\
Postboks 7803, N-5020  Bergen, Norway\\
\end{center}

\vskip 2cm

\begin{abstract}

CP violation plays a very important role in nature with implications both for Particle Physics and for Cosmology. Accounting for the observed  matter--antimatter asymmetry of the Universe requires the existence of new sources of CP violation beyond the Standard Model. In models with an extended scalar sector CP violation can emerge either explicitly, i.e., at the Lagrangian level, or spontaneously.  Spontaneous CP violation occurs in the framework of the electroweak  symmetry breaking whenever the Lagrangian conserves CP and the vacuum breaks it. This requires that not all vacuum expectation values be real. In the context of multi-Higgs extensions of the Standard Model imposing the existence of a scalar basis where all couplings  are real is a sufficient condition for CP to be explicitly conserved. We discuss a three-Higgs-doublet model with an underlying $S_3$ symmetry, allowing in principle for complex couplings. In this framework it is possible to have either spontaneous or explicit CP violation in the scalar sector, depending on the regions of parameter space corresponding to the different possible vacua of the $S_3$ symmetric potential. We list all possible vacuum structures allowing for CP violation in the scalar sector specifying whether it can be explicit or spontaneous. It is by now established that CP is violated in the flavour sector and that the Cabibbo-Kobayashi-Maskawa matrix is complex. In order to understand what are the possible sources of CP violation in the Yukawa sector we analyse the implications of the different available choices of representations for the quarks under the $S_3$ group. This classification is based strictly on the exact $S_3$-symmetric scalar potential with no soft symmetry breaking terms. The scalar sector of one such model was explored numerically. After applying the theoretical and the most important experimental constraints the available parameter space is shown to be able to give rise to light neutral scalars at the $\mathcal{O}(\text{MeV})$ scale.

\end{abstract}

\end{titlepage}

\section{Introduction}

It is by now established that the Standard Model (SM) of Particle Physics cannot be the final theory. In fact, on the one hand the SM leaves several questions open and on the other hand there is already clear evidence for physics beyond the SM. The phenomenon of neutrino oscillations requires the extension of the leptonic sector of the SM. Accounting for the observed  baryon asymmetry of the Universe requires new sources of CP violation~\cite{Sakharov:1967dj,Kuzmin:1985mm,Gavela:1993ts,Huet:1994jb,Gavela:1994dt}. Furthermore, there are by now several experimental anomalies in the flavour sector hinting at the existence of new physics~\cite{Graverini:2018riw,Belle-II:2018jsg}. 

There are strong motivations to consider multi-Higgs extensions of the SM, despite the fact that the properties of the Higgs boson discovered in 2012~\cite{ATLAS:2012yve,CMS:2012qbp,ATLAS:2015yey,ATLAS:2016neq} are still in experimental agreement with the SM Higgs predictions ~\cite{CMS:2022dwd,ATLAS:2022vkf}. Among these motivations are the possibility of having new sources of CP violation coming from this sector as well as having CP spontaneously broken in the Lagrangian, rather than explicitly through the introduction of complex Yukawa couplings. In the case of spontaneous CP breaking the complex parameters of the Cabibbo–Kobayashi–Maskawa (CKM) matrix would be generated through phases of the vacuum expectation values (vev) of Higgs fields.
 
Among the simplest extensions of the SM are the so-called two-Higgs-doublet models (2HDM)~\cite{Gunion:1989we,Branco:2011iw}, in which a second SU(2) scalar doublet with the same quantum numbers as the one of the SM is introduced, allowing for explicit CP violation in the scalar sector. It was shown, long ago~\cite{Lee:1973iz}, that the  2HDM also allows for spontaneous CP violation. However, the 2HDM generates potentially dangerous flavour changing neutral currents (FCNC) at tree level, when the fermions are allowed to freely couple to both Higgs bosons. There are stringent experimental limits on FCNC, thus requiring the introduction of symmetries to control such effects in multi-Higgs models. It is possible to control FCNC in a natural way by introducing a  $\mathbb{Z}_2$ symmetry,~\cite{Glashow:1976nt,Paschos:1976ay} under which some of the Higgs doublets and the quark fields transform non-trivially, leading to natural flavour conservation (NFC). However, the 2HDM with an imposed  (exact) $\mathbb{Z}_2$ symmetry no longer provides additional sources of CP violation. Such possibility can be restored by introducing a soft breaking term (violation of the symmetry through an operator of dimension 2) of the $\mathbb{Z}_2$ symmetry~\cite{Branco:1985aq}. The CP properties and cosmological implications of the 2HDM have been extensively studied and are still of great interest~\cite{Branco:1999fs,Gunion:1989we,Branco:2011iw}. 

Despite the rich phenomenology of the 2HDM, models with three Higgs doublets are attracting a lot of interest in the literature. Within three-Higgs-doublet models (3HDM) it is possible to have an additional source of CP violation in the scalar sector while at the same time having natural flavour conservation~\cite{Weinberg:1976hu,Deshpande:1976yp,Gatto:1979mr,Branco:1979pv,Branco:1980sz}. The predictability of models with several SU(2) scalar doublets can be quickly lost due to the rapid growth of free parameters as the number of doublets increases~\cite{Olaussen:2010aq}. Therefore, it is essential to control the number of free parameters by means of symmetries~\cite{Ishimori:2010au,Ivanov:2012fp,Ivanov:2014doa}. It was found that in many cases imposing additional symmetries to multi-Higgs models eliminates the possibility of having CP violation in the Higgs sector. Many examples of different symmetries have been studied where CP can be violated in the scalar sector, in the context of 3HDM, including $A_4$ and $S_4$~\cite{deAdelhartToorop:2010jxh,Degee:2012sk,GonzalezFelipe:2013xok}, $\Delta(27)$~\cite{Branco:1983tn,deMedeirosVarzielas:2011zw,deMedeirosVarzielas:2012ylr,Ma:2013xqa,Nishi:2013jqa,Fallbacher:2015rea}, exotic CP4~\cite{Ivanov:2015mwl,Ferreira:2017tvy,Ivanov:2018ime,Ivanov:2021pnr}. Unlike in the case of the 2HDM (Inert Doublet Models~\cite{Barbieri:2006dq,LopezHonorez:2006gr,Cao:2007rm}), it is possible to have both CP violation (in the active or inert sector) and a possible dark matter (DM) candidate in the context of 3HDM~\cite{Grzadkowski:2009bt,Grzadkowski:2010au,Cordero-Cid:2016krd,Azevedo:2018fmj,Cordero-Cid:2018man,Ivanov:2018srm,Kuncinas:2022whn,Biermann:2022meg}.

The $S_3$-symmetric 3HDM has been studied since the late 70s in the irreducible~\cite{Pakvasa:1977in} and reducible~\cite{Derman:1978rx,Derman:1979nf} representations, trying to explain the fermionic structures. 
Since then, different cases of the CP-conserving scalar potential were discussed~\cite{Machado:2012gxi,Chakrabarty:2015kmt,Gomez-Bock:2021uyu,Khater:2021wcx}. With real couplings CP is explicitly conserved by the scalar potential and there is only the possibility of having spontaneous CP violation for special vacua~\cite{Emmanuel-Costa:2016vej}. In this paper we study the CP properties of the $S_3$-symmetric 3HDM allowing for complex couplings in the potential. Different vacua correspond to different regions of parameter space which are determined by imposing the minimisation conditions.  In the case of complex couplings some of the CP-conserving vacuum structures of the real potential now correspond to regions of parameter space that allow for explicit CP violation. For instance, CP can be explicitly violated even in the case where the vacuum preserves the $S_3$ symmetry.

The paper is organised as follows. In Section~\ref{Sec:Scalar_potential} we introduce the $S_3$-symmetric scalar potential along with a discussion of the starting point for our analysis corresponding to a suitable choice of a scalar basis. Although two new phases can be introduced in the quartic couplings, we choose to fix one of them to be zero and we allow for the vevs to be complex, which is required for generality. After fixing the basis for our discussion, in Section~\ref{Sec:CP_invariants} we employ the powerful tool of CP-odd Higgs basis invariants in order to obtain constraints on the parameter space for the case of explicit CP conservation. The CP-odd basis invariant conditions require the imaginary part of different possible combinations of $Y$- and $Z$-tensors to vanish. With this information it is then possible to classify models based on the structure of their vevs. This is done in Section~\ref{Sec:Different_Models}. In Section~\ref{Sec:Yukawa_Lagrangian} we cover the building of the Yukawa Lagrangian assuming that fermions are charged under the $S_3$ symmetry. Some of the models require further numerical investigation. These models are discussed in Section~\ref{Sec:Num_exploration}. In section~\ref{Sec:Discussion} we present our conclusions.

\section{The scalar potential}\label{Sec:Scalar_potential}

The $S_3$-symmetric 3HDM is in the irreducible representation given by a singlet, a pseudosinglet and a doublet. We shall adopt the singlet-doublet representation. In this case the $S_3$-symmetric potential can be written as~\cite{Kubo:2004ps,Teshima:2012cg,Das:2014fea}:
\begin{subequations}\label{Eq:Vscalar}
\begin{align}
V_2 =&\,\mu_0^2 h_S^\dagger h_S +\mu_1^2(h_1^\dagger h_1 + h_2^\dagger h_2), \label{Eq:Vscalar_V2} \\
\begin{split} V_4 =&\,
\lambda_1(h_1^\dagger h_1 + h_2^\dagger h_2)^2 
+\lambda_2(h_1^\dagger h_2 - h_2^\dagger h_1)^2
+\lambda_3[(h_1^\dagger h_1 - h_2^\dagger h_2)^2+(h_1^\dagger h_2 + h_2^\dagger h_1)^2] \\
&+ \left\lbrace\lambda_4\left[(h_S^\dagger h_1)(h_1^\dagger h_2+h_2^\dagger h_1)+(h_S^\dagger h_2)(h_1^\dagger h_1-h_2^\dagger h_2) \right]+\hc\right\rbrace \\
&+\lambda_5(h_S^\dagger h_S)(h_1^\dagger h_1 + h_2^\dagger h_2) +\lambda_6[(h_S^\dagger h_1)(h_1^\dagger h_S)+(h_S^\dagger h_2)(h_2^\dagger h_S)] \\
&+ \left\lbrace\lambda_7 \left[(h_S^\dagger h_1)(h_S^\dagger h_1) + (h_S^\dagger h_2)(h_S^\dagger h_2)\right] +\hc\right\rbrace+\lambda_8(h_S^\dagger h_S)^2.\end{split}
\end{align}
\end{subequations}
There are two couplings, $\lambda_4$ and $\lambda_7$, that could be complex. Hence, CP symmetry can be broken explicitly. All other couplings have to be real due to the hermiticity of the potential. 

Another option would be to consider the pseudosinglet-doublet representation. In this case there is no unitary transformation into the defining representation of $S_3$. Such representation would yield an equivalent scalar potential, however the Yukawa Lagrangian would not be equivalent. We do not consider such representation here.

The $S_3$ singlet and doublet fields will be decomposed as
\begin{equation} \label{Eq.Doublets_components}
h_S =\left(
\begin{array}{c}h_S^+\\ (w_S+ \eta_S+i \chi_S)/\sqrt{2}
\end{array}\right),\qquad
h_i =\left(
\begin{array}{c}h_i^+\\ (w_i+\eta_i+i \chi_i)/\sqrt{2}
\end{array}\right), \quad i=\{1,2\},
\end{equation}
where the $w_S$ and $w_i$ parameters can be complex. If CP is not broken explicitly, such complex vevs could result in spontaneous CP violation. The complex vevs shall be written as
\begin{equation*}
\{w_S,\, w_1,\, w_2 \} \to \{\hat w_S, \, \hat w_1 e^{i \sigma_1},\, \hat w_2 e^{i \sigma_2}\},
\end{equation*}
following the notation of Ref.~\cite{Emmanuel-Costa:2016vej}. The hatted $w$s represent absolute values. Due to the overall U(1) symmetry of the scalar potential it is possible to rotate away one of the phases, which was chosen to be that of the $S_3$ singlet, $\hat w_S$.

A different approach would be to write the scalar potential in terms of a reducible triplet, as done by Derman~\cite{Derman:1978rx}. He expressed the potential as a sum of five expressions, each paired with its hermitian conjugate. However, due to the underlying $S_3$ symmetry, not all the hermitian conjugate pairs can be accompanied by a complex coefficient. Transformations between the reducible and irreducible basis are covered in appendix~\ref{App:Derman_pot}.

\subsection{Possible choices of complex coefficients}

The $S_3$-symmetric 3HDMs were classified in Ref.~\cite{Emmanuel-Costa:2016vej}. Here, we are interested in expanding the set of solutions by considering complex couplings. For the purpose of the current work we need to define a suitable basis for the scalar potential. The most general approach would rely on the fact that both, or either, of $\{\lambda_4,\, \lambda_7\} \in \mathbb{C}$. However, such an approach would yield redundant solutions (models). In other words, there would exist different descriptions of one and the same physical situation, related by a unitary transformation. As we shall show, it is sufficient to consider instances when either $\lambda_4$ or $\lambda_7$ acquires a non-vanishing phase. We shall see that for the purpose of discussing CP-conserving limits of the potential, it is convenient to take $\lambda_4$ complex and $\lambda_7$ real.

When dealing with complex potential parameters it might be useful to perform a basis change so that some of the phases are rotated away. Let us consider the following basis rotation of two of the SU(2) doublets,
\begin{equation}\label{Eq.h_ph_theta}
h_i = e^{i \theta_i} h_i^\prime, \quad i=\{1,2\}.
\end{equation}
Due to the global U(1) symmetry the phase of the $S_3$ singlet, $h_S$, can always be rotated away, hence it is not considered. In total, there are four couplings sensitive to such rotations: $\{ \lambda_2+\lambda_3,\, \lambda_4,\,\lambda_7\}$. As noted earlier, in the generic singlet-doublet representation basis only two coefficients could have a phase, $\{\lambda_4,\,\lambda_7\}$. In consistency with the basis change of eq.~\eqref{Eq.h_ph_theta} we shall write couplings in a complex polar notation
\begin{equation}\label{Eq.l4l7_alpha}
\lambda_i = e^{i \alpha_i}|\lambda_i|, \quad i=\{4,7\}. 
\end{equation}
At this point, we have introduced three sets of phases: the $\sigma_i$ vev phases, $\theta_i$ describing a basis change, and $\alpha_i$ describing a polar rotation of $\lambda_4$ and $\lambda_7$. For simplicity we shall drop primes from the factors appearing in a new basis.

After rotating the SU(2) doublets according to eq.~(\ref{Eq.h_ph_theta}), the quartic part of the scalar potential can be split into two parts,
\begin{equation}
V_4=V_4^0+V_4^\text{ph},
\end{equation}
where the phase insensitive, $V_4^0$, and the phase sensitive, $V_4^\text{ph}$, quartic potential parts are:
\newpage
\begin{subequations}
\begin{align}
\begin{split}V_4^0 &= \lambda_1\left(h_1^\dagger h_1 + h_2^\dagger h_2\right)^2 - 2 \lambda_2 \left(h_1^\dagger h_2\right)\left(h_2^\dagger h_1\right)\\
&\quad+\lambda_3\left[\left(h_1^\dagger h_1 - h_2^\dagger h_2\right)^2 + 2 \left(h_1^\dagger h_2\right)\left(h_2^\dagger h_1\right)\right]+\lambda_5\left(h_S^\dagger h_S\right)\left(h_1^\dagger h_1 + h_2^\dagger h_2\right)\\
&\quad+\lambda_6\left[\left(h_S^\dagger h_1\right)\left(h_1^\dagger h_S\right)+\left(h_S^\dagger h_2\right)\left(h_2^\dagger h_S\right)\right]+\lambda_8\left(h_S^\dagger h_S\right)^2,\end{split}\\
\begin{split}  V_4^\text{ph} &= 
e^{-2i(\theta_1 - \theta_2)} \left( \lambda_2 + \lambda_3 \right) \left(h_1^\dagger h_2\right)^2\\
&\quad + |\lambda_4| \bigg\{ e^{i(2 \theta_1 - \theta_2 + \alpha_4)} \left(h_S^\dagger h_1\right)\left(h_2^\dagger h_1\right) \\
&\qquad\qquad + e^{i(\theta_2 + \alpha_4)}\left[ \left(h_S^\dagger h_1\right)\left(h_1^\dagger h_2\right) + \left(h_S^\dagger h_2\right)\left( h_1^\dagger h_1 - h_2^\dagger h_2\right)  \right]  \bigg\} \\
&\quad + |\lambda_7| \left[e^{i(2 \theta_1 + \alpha_7)} \left(h_S^\dagger h_1\right)^2 + e^{i(2 \theta_2 + \alpha_7)} \left(h_S^\dagger h_2\right)^2\right] + \hc
\end{split}\label{Eq:V4C_explicit_phases-0}
\end{align}
\end{subequations}
Here, we see that the scalar potential is sensitive to different $\theta_i$ phases of eq.~\eqref{Eq.h_ph_theta}. In principle, one can consider a basis with real vevs and complex couplings, as presented above. We note that the sum $(\lambda_2 + \lambda_3)$ would then get a phase, while the form of the scalar potential~\eqref{Eq:Vscalar} suggests that these couplings should be real. This is due to a possible choice of a basis. In order to simplify the book-keeping, without loss of generality, we choose $\theta_1=\theta_2\equiv \theta$ (and allow for complex vevs). Thus, $(\lambda_2 + \lambda_3)$ remains real and $\lambda_4$ and $\lambda_7$ each change by an overall phase rotation; meaning that either one of them can be made real. Explicitly, this choice gives for the phase-dependent part,
\begin{align} \label{Eq:V4C_explicit_phases}
\left(V_4^\text{ph}\right)^\prime ={}& |\lambda_4| e^{i(\theta + \alpha_4)} \bigg\{ \left(h_S^\dagger h_1\right)\left(h_2^\dagger h_1\right) 
+ \left(h_S^\dagger h_1\right)\left(h_1^\dagger h_2\right) + \left(h_S^\dagger h_2\right)\left( h_1^\dagger h_1 - h_2^\dagger h_2\right)  \bigg\}\nonumber\\ 
&+|\lambda_7| e^{i(2 \theta + \alpha_7)} \left[\left(h_S^\dagger h_1\right)^2 + \left(h_S^\dagger h_2\right)^2\right] + \hc
\end{align}

Furthermore, we may rotate one of the $\alpha_i$ phases away. We would like to stress that we are only interested in cases with non-vanishing phases of the couplings. Cases with spontaneous CP violation, and strictly real couplings, were covered in Ref.~\cite{Emmanuel-Costa:2016vej}. 

After checking the minimisation conditions of all models (presented in Section~\ref{Sec:Different_Models}), we concluded that it is convenient to choose a basis where the $\lambda_7$ coupling becomes real, $2 \theta + \alpha_7 = 0 \mod \pi$. For simplicity, we shall assume that $\lambda_4$ of eq.~\eqref{Eq:Vscalar} is split into a real and an imaginary part
\begin{equation}\label{Eq:lambda_4_explicit_l4R+l4I}
\lambda_4 \xrightarrow[]{\eqref{Eq.l4l7_alpha}} e^{i \alpha_4}|\lambda_4| \equiv \lambda_4^\mathrm{R} + i \lambda_4^\mathrm{I}.
\end{equation}
Furthermore, since the scalar potential is not invariant under a phase rotation~\eqref{Eq.h_ph_theta} we shall consider both real and complex vacua.

\section{Explicit CP violation}\label{Sec:CP_invariants}
Allowing for complex parameters in the potential of the $S_3$-symmetric 3HDM opens up the possibility for explicit CP violation. If the potential explicitly violates CP, the CP violation will either be hard (CP violating phases present in $V_4$ cannot be removed from $V_4$ by basis changes) or soft (CP violating phases present in the potential can by a change of basis be moved to $V_2$, but they cannot be removed from the potential). It has been shown that there exist models where phases present in the potential cannot be removed from the potential using basis changes, yet the potential is CP invariant, i.e. CP4 \cite{Ivanov:2015mwl,Ferreira:2017tvy,Ivanov:2018ime,Ivanov:2021pnr}. When we refer to irremovable CP violating phases, we do not refer to the irremovable phases of CP4 models since such phases do not violate CP.
If the potential does not explicitly violate CP, the nature of the vacuum will determine the CP properties of the model, leaving open the possibilities of spontaneous CP violation or CP conservation. We shall proceed to write the scalar potential~\eqref{Eq:Vscalar} in a more compact form,
\begin{subequations}
	\begin{align}
		V_2={}&Y_{ab} \left( h^\dagger_a h_b \right),\\
		V_4={}&\frac{1}{2}Z_{abcd}\left(h^\dagger_a h_b\right)\left(h^\dagger_c h_d\right),
	\end{align}
\end{subequations}
where the non-zero elements of the $Y$- and $Z$-tensors are
\begin{equation}
	\begin{aligned}
		&\begin{split}
			&Y_{11}=Y_{22}=\mu_1^2,\\
			&Z_{1111}=Z_{2222}=2\lambda_1+2\lambda_3,\\
			&Z_{1122}=Z_{2211}=2\lambda_1-2\lambda_3,\\
			&Z_{1221}=Z_{2112}=-2\lambda_2+2\lambda_3,\\
			&Z_{1212}=Z_{2121}=2\lambda_2+2\lambda_3,\\
		\end{split}\hspace{20pt}
		\begin{split}
			&Y_{33}=\mu_0^2,\\
			&Z_{3333}=2\lambda_8,\\
			&Z_{1133}=Z_{2233}=Z_{3311}=Z_{3322}=\lambda_5,\\
			&Z_{1331}=Z_{2332}=Z_{3113}=Z_{3223}=\lambda_6,\\
			&Z_{1313}=Z_{2323}=Z_{3131}=Z_{3232}=2\lambda_7,\\
		\end{split}\\
		&Z_{1123}=Z_{1213}=Z_{1312}=Z_{1321}=Z_{2113}=Z_{2311}=-Z_{2223}=-Z_{2322}=\lambda_4^\mathrm{R} - i \lambda_4^\mathrm{I},\\ &Z_{1132}=Z_{1231}=Z_{2131}=Z_{3112}=Z_{3121}=Z_{3211}=-Z_{2232}=-Z_{3222}=\lambda_4^\mathrm{R} + i \lambda_4^\mathrm{I}.
	\end{aligned}
\end{equation}
This form enables us to easily establish quantities that are invariant under basis changes. Utilizing the elegant technique of creating CP-odd invariants from the $Y$- and $Z$-tensors will provide us with a powerful and elegant tool for establishing the CP properties of the potential. For descriptions of this technique consult Refs.~\cite{Mendez:1991gp,Lavoura:1994fv,Botella:1994cs,Branco:2005em,Davidson:2005cw,Gunion:2005ja,Haber:2006ue}. Some other methods were discussed in Refs.~\cite{Nishi:2006tg,deMedeirosVarzielas:2016rii,Ogreid:2017alh,Trautner:2018ipq,Ivanov:2019kyh}.
We shall start by defining six CP-odd invariants, namely
\begin{subequations}\label{Eq:I6_inv}
\begin{align}
\mathrm{I}_{5Z}^{(1)}={}& \mathbb{I}\mathrm{m} \left[Z_{aabc}Z_{dbef}Z_{cghe}Z_{idgh}Z_{fijj}\right],\\
\mathrm{I}_{5Z}^{(2)}={}& \mathbb{I}\mathrm{m} \left[Z_{abbc}Z_{daef}Z_{cghe}Z_{idgh}Z_{fjji}\right],\\
\mathrm{I}_{6Z}^{(1)}={}& \mathbb{I}\mathrm{m}\left[Z_{abcd}Z_{baef}Z_{gchi}Z_{djke}Z_{fkil}Z_{jglh}\right],\\
\mathrm{I}_{6Z}^{(2)}={}& \mathbb{I}\mathrm{m}\left[Z_{abcd}Z_{baef}Z_{gchi}Z_{dejk}Z_{fhkl}Z_{lgij}\right],\\
\mathrm{I}_{7Z}={}& \mathbb{I}\mathrm{m}\left[Z_{abcd}Z_{eafc}Z_{bgdh}Z_{iejk}Z_{gflm}Z_{hlkn}Z_{minj}\right],\\
\stepcounter{parentequation}
\gdef\theparentequation{\sillymacro}
\setcounter{equation}{-1}
\mathrm{I}_{2Y3Z}={}& \mathbb{I}\mathrm{m} \left[Z_{abcd}Z_{befg}Z_{dchf}Y_{ga}Y_{eh} \right] \label{Eq:I2Y3Z_inv}.
\end{align}
\end{subequations}
Next, we establish two theorems which serve as invaluable tools in the discussion of the CP properties of the potential.

\begin{theorem}
	The quadrilinear part of the $S_3$-symmetric 3HDM potential, $V_4$, explicitly conserves CP if and only if $\mathrm{I}_{5Z}^{(1)}=\mathrm{I}_{5Z}^{(2)}=\mathrm{I}_{6Z}^{(1)}=\mathrm{I}_{6Z}^{(2)}=\mathrm{I}_{7Z}=0$. 
\end{theorem}
\begin{proof}
	We must prove that the two statements of the theorem imply each other. First, we prove that if $V_4$ is CP invariant, then $\mathrm{I}_{5Z}^{(1)}=\mathrm{I}_{5Z}^{(2)}=\mathrm{I}_{6Z}^{(1)}=\mathrm{I}_{6Z}^{(2)}=\mathrm{I}_{7Z}=0$. This follows by definition, since whenever $V_4$ is CP invariant, all CP-odd invariants constructed from only $Z$-tensors must vanish.
	
	Next, we must prove that $\mathrm{I}_{5Z}^{(1)}=\mathrm{I}_{5Z}^{(2)}=\mathrm{I}_{6Z}^{(1)}=\mathrm{I}_{6Z}^{(2)}=\mathrm{I}_{7Z}=0$ imply a CP invariant $V_4$. In order to do this, we demand the vanishing of all five invariants and work out algebraically solutions in terms of the potential parameters:
	\begin{itemize}
		\item {\bf Solution 0:} $\lambda_4^\mathrm{I}=0$;
		\item {\bf Solution 1:} $\lambda_4^\mathrm{R}=0$;
		\item {\bf Solution 2:} $\lambda_7=0$;
		\item {\bf Solution 3 (\boldmath$\lambda_4^\mathrm{R} \lambda_4^\mathrm{I} \lambda_7  \neq  0$): }
		\begin{gather*}
			\begin{aligned}
				\left( \lambda_4^\mathrm{R} \right)^2={}& -\frac{(\lambda_{23}-\lambda_7)(2\lambda_{23}+\lambda_7)^2}{\lambda_7},\\
				\left( \lambda_4^\mathrm{I} \right)^2={}& \frac{(\lambda_{23}+\lambda_7)(2\lambda_{23}-\lambda_7)^2}{\lambda_7},
			\end{aligned}\hspace{50pt}
			\begin{aligned}
				\lambda_5 ={}& 2 \left( \lambda_1 + \lambda_2 \right),\\
				\lambda_6 ={}& 4 \lambda_3,\\
				\lambda_8 ={}& \lambda_1 - \lambda_2.
			\end{aligned}
		\end{gather*}
		\end{itemize}
For each of these solutions we were able to show that there exists a real basis for $V_4$, which concludes the proof of the theorem. The technical details demonstrating this are relegated to Appendix~\ref{App:CPodd_inv}.	
\end{proof}

This theorem deals only with properties of $V_4$ and gives us the conditions for CP invariance of $V_4$. The possibility of having explicit CP violation is still present, but CP must then be softly broken (CP violating phases can be transferred to $V_2$ by a change of basis). As shown in Appendix~\ref{App:CPodd_inv}, Solutions 0--2 do not allow for explicit CP violation, only spontaneous CP violation is possible. The possibility of explicit CP violation is therefore restricted to Solution 3. For Solution 3, the basis transformations to a real $V_4$ basis will, in general, generate extra terms in the transformed $V_2$, containing also complex parameters. 

One might wonder if there could exist an explicitly CP conserving $S_3$-symmetric 3HDM without the existence of a real basis, like in CP4 models. The proof of the theorem tells us that this is not the case for the complex $S_3$-symmetric 3HDM. We show in Appendix~\ref{App:CPodd_inv} that all the possible solutions can be written in a real basis when all of the five invariants vanish. In conclusion, a solution without a real basis does not exist.

Finally, we also include the quadratic part of the potential, $V_2$, and formulate the CP properties of the whole potential in our second theorem.
\begin{theorem}
	The $S_3$-symmetric 3HDM potential, $V=V_2+V_4$, explicitly conserves CP if and only if $\mathrm{I}_{5Z}^{(1)}=\mathrm{I}_{5Z}^{(2)}=\mathrm{I}_{6Z}^{(1)}=\mathrm{I}_{6Z}^{(2)}=\mathrm{I}_{7Z}=\mathrm{I}_{2Y3Z}=0$.
\end{theorem}
\begin{proof}
	Again, we must prove that the two statements of the theorem imply each other. First, we prove that if $V$ is CP invariant, then $\mathrm{I}_{5Z}^{(1)}=\mathrm{I}_{5Z}^{(2)}=\mathrm{I}_{6Z}^{(1)}=\mathrm{I}_{6Z}^{(2)}=\mathrm{I}_{7Z}=\mathrm{I}_{2Y3Z}=0$. This follows by definition, since whenever the scalar potential, $V$,  is CP invariant, all CP-odd invariants constructed from only $Y$- and  $Z$-tensors must vanish.
	
	Then, we must prove that $\mathrm{I}_{5Z}^{(1)}=\mathrm{I}_{5Z}^{(2)}=\mathrm{I}_{6Z}^{(1)}=\mathrm{I}_{6Z}^{(2)}=\mathrm{I}_{7Z}=\mathrm{I}_{2Y3Z}=\mathrm{I}_{2Y3Z}=0$ implies a CP invariant $V$. In order to do this we demand the vanishing of all six invariants and work out algebraically the solutions in terms of the potential parameters. The difference from Theorem 1 is that we now include the vanishing of $\mathrm{I}_{2Y3Z}$. We find that Solutions~0--2 make all six invariants vanish. On the other hand, Solution 3 is modified with one additional constraint, namely \mbox{$\mu_1^2=\mu_0^2$}:
\begin{itemize}
\item {\bf Solution 3$^\prime$ (\boldmath$\lambda_4^\mathrm{R} \lambda_4^\mathrm{I} \lambda_7  \neq  0$): }
\begin{gather*}
\begin{aligned}
\mu_1^2 ={}& \mu_0^2,\\
\left( \lambda_4^\mathrm{R} \right)^2={}& -\frac{(\lambda_{23}-\lambda_7)(2\lambda_{23}+\lambda_7)^2}{\lambda_7},\\
\left( \lambda_4^\mathrm{I} \right)^2={}& \frac{(\lambda_{23}+\lambda_7)(2\lambda_{23}-\lambda_7)^2}{\lambda_7},
\end{aligned}\hspace{50pt}
\begin{aligned}
&\\
\lambda_5 ={}& 2 \left( \lambda_1 + \lambda_2 \right),\\
\lambda_6 ={}& 4 \lambda_3,\\
\lambda_8 ={}& \lambda_1 - \lambda_2.
\end{aligned}
\end{gather*}
\end{itemize}
For Solutions 0--2, the changes of basis to a real $V_4$ basis were expressed in terms of a phase rotation of a doublet, and did not affect $V_2$, which stayed real. For Solution 3$^\prime$, the constraint $\mu_1^2=\mu_0^2$ yields $V_2$ invariant under the U(3) transformations, and hence the quadratic $V_2$ remains real. To summarise, for each of the solutions found we were able to show that there exists a real basis for $V$, and this concludes the proof of the theorem.	
\end{proof}

This theorem deals only with the CP properties of the potential, not the vacuum. CP violation may still occur, but CP will then be spontaneously broken by the vacuum.

We emphasise that the two theorems only apply to the $S_3$-symmetric 3HDM. For the general 3HDM, the necessary and sufficient set of the CP-odd invariants needed for explicit CP conservation has not yet been identified. The above theorems are formulated in terms of basis invariant quantities, hence they may be used to determine the CP properties of the potential in any scalar basis. 

We are now in a position to discuss and classify different cases of explicit CP conservation of the $S_3$-symmetric 3HDM, based on these theorems. There is a common factor, $\lambda_4^\mathrm{R} \lambda_4^\mathrm{I} \lambda_7$, contained within every CP-odd $\mathrm{I}$-invariant.

For Solutions 0--2 it suffices to have  $\lambda_4^\mathrm{R} \lambda_4^\mathrm{I} \lambda_7=0$ to eliminate the possibility of having explicit CP violation. Solution 0, given by $\lambda_4^\mathrm{I}=0$, reduces to the cases studied in Ref.~\cite{Emmanuel-Costa:2016vej}. For Solutions 1 and 2 it is easy to find simple basis changes that make all potential parameters real as given in Appendix \ref{App:CPodd_inv}.

Solution 3 requires some further explanation. No additional continuous symmetries are realised in the scalar potential when the CP-conserving conditions of eqs.~\eqref{Eq:Sol_CP_Z7} are applied. This was verified by checking the scalar mass eigenstates and the claim was also confirmed by going through the basis-independent conditions provided in Ref.~\cite{deMedeirosVarzielas:2019rrp}. However, after applying the conditions for CP-odd invariants to vanish the $S_3$-symmetric potential gets enlarged to $\Delta(54)$. This can be verified by performing a basis rotation
\begin{equation}
\begin{pmatrix}
h_1 \\
h_2 \\
h_S
\end{pmatrix} = \frac{1}{\sqrt{2}}\begin{pmatrix}
1 & -i & 0 \\
-i & 1 & 0 \\
0 & 0 & \sqrt{2}
\end{pmatrix} \begin{pmatrix}
\phi_1 \\
\phi_2 \\
\phi_3 
\end{pmatrix}.
\end{equation}

In the new basis the scalar potential becomes
\begin{subequations}\label{Eq:Delta54_Pot}
\begin{align}
V_2  ={}& \mu_1^2(\phi_1^\dagger \phi_1 + \phi_2^\dagger \phi_2 + \phi_3^\dagger \phi_3),\\
\begin{split}
V_4 ={}& \frac{1}{3}\left( 3 \lambda_1 + \lambda_2 \right)\left(\phi_1^{\dagger} \phi_1+\phi_2^{\dagger} \phi_2+\phi_3^{\dagger} \phi_3\right)^2 +4\lambda_3\left(\left|\phi_1^{\dagger} \phi_2\right|^2+\left|\phi_2^{\dagger} \phi_3\right|^2+\left|\phi_3^{\dagger} \phi_1\right|^2\right)  \\
 & - \frac{4 \lambda_2}{3}\left[\left(\phi_1^{\dagger} \phi_1\right)^2+\left(\phi_2^{\dagger} \phi_2\right)^2+\left(\phi_3^{\dagger} \phi_3\right)^2 \right.\\
 &\hspace{86pt}\left.-\left(\phi_1^{\dagger} \phi_1\right)\left(\phi_2^{\dagger} \phi_2\right)-\left(\phi_2^{\dagger} \phi_2\right)\left(\phi_3^{\dagger} \phi_3\right)-\left(\phi_3^{\dagger} \phi_3\right)\left(\phi_1^{\dagger} \phi_1\right)\right]\\
&+\left\{ 2i \lambda_7 \left(\phi_1^{\dagger} \phi_3\right)\left(\phi_2^{\dagger} \phi_3\right)
+ \sqrt{2}\lambda_4\left(\phi_2^{\dagger} \phi_1\right)\left(\phi_3^{\dagger} \phi_1\right)
- i\sqrt{2}\lambda_4\left(\phi_3^{\dagger} \phi_2\right)\left(\phi_1^{\dagger} \phi_2\right) + \mathrm{h.c.} \right\},
\end{split} \raisetag{3.5\baselineskip}
\end{align}
\end{subequations}
where $\mu_0^2 = \mu_1^2$ was imposed as required by eq.~\eqref{Eq:Mu_Eq_Req}, and where $\lambda_7$ is real and $\lambda_4$ is complex, see eq.~\eqref{Eq:lambda_4_explicit_l4R+l4I}. The potential has the structure of the $\Delta(54)$-symmetric one, as given by eqs.~(52) and (53) in Ref.~\cite{deMedeirosVarzielas:2019rrp}. 

\section{CP violation in different vacua}\label{Sec:Different_Models}

We classify cases with complex scalar potential based on Ref.~\cite{Emmanuel-Costa:2016vej}. We first list cases allowing for spontaneous CP violation when the scalar potential is real:
\begin{itemize}
\item C-III-a $(0,\,\hat w_2 e^{i \sigma_2},\, \hat w_S)$;
\item C-III-h $(\sqrt{3} \hat w_2 e^{i \sigma_2},\, \pm \hat w_2 e^{i \sigma_2} ,\, \hat w_S)$;
\item C-IV-c $\left(\sqrt{1 + 2 \cos^2 \sigma_2} \hat w_2 ,\, \hat w_2 e^{i \sigma_2} ,\, \hat w_S\right)$;
\item C-IV-f $\left(\sqrt{2 + \frac{\cos(\sigma_1 - 2\sigma_2)}{\cos \sigma_1}} \hat w_2 e^{i \sigma_1},\, \hat w_2 e^{i \sigma_2} ,\, \hat w_S\right)$;
\end{itemize}

\subsection{Real vacua}\label{Sec:Diff_Re_Models}

We consider real vacua with $\lambda_7$ real and $\lambda_4$ complex. In this case it is possible to have explicit CP violation. 
The minimisation conditions are provided in appendix~\ref{App:Derivatives_Rvev}. In some cases the minimisation conditions require $\lambda_4^\mathrm{I}=0$. Therefore, we do not consider such models, to wit:
\begin{itemize}
\item R-II-1a $(0,\, w_2,\, \hat w_S)$;
\item R-II-1b,c $(\mp \sqrt{3} w_2,\, w_2,\, \hat w_S)$;
\item R-II-2 $(0,\, w_2,\, 0)$;
\item R-II-3 $(w_1,\, w_2,\, 0)$;
\item R-III $(w_1,\, w_2,\, \hat w_S)$;
\end{itemize}

By checking the minimisation conditions \eqref{Eq:DV_Dchi1}--\eqref{Eq:DV_Dchi3} we can identify which models do not require $\lambda_4^\mathrm{I} = 0$. In neither of the models listed below did we find instances when the CP-odd invariants would vanish. Therefore, all these models are CP violating in the scalar sector:

\begin{itemize}

\item  \textbf{R-I-1: \boldmath$(0,\,0,\,\hat w_S)$}\\
In this model there is a single minimisation condition given by
\begin{equation}
\mu_0^2=-\hat w_S^2\lambda_8.
\end{equation}
There is a pair of charged mass-degenerate states and two pairs of neutral mass-degenerate states.

\item \textbf{R-I-2a: \boldmath$(w_1,\, 0,\, 0)$}\\
In this model there is a single minimisation condition given by
\begin{equation}
\mu_1^2=-(\lambda_1+\lambda_3)w_1^2.
\end{equation}

\item \textbf{R-I-2b,c: \boldmath$( w_1,\, \pm \sqrt{3} w_1,\, 0)$}\\
In this model there is a single minimisation condition given by
\begin{equation}
\mu_1^2=-4(\lambda_1+\lambda_3)w_1^2.
\end{equation}
\end{itemize}

\subsection{Complex vacua}\label{Sec:Diff_Im_Models}

Next, we cover cases with the vacuum form given by $\{\hat w_1 e^{i \sigma_1},\, \hat w_2 e^{i \sigma_2}, \, \hat w_S\}$. The minimisation conditions are provided in appendix~\ref{App:Derivatives_Cvev}. As before, in some cases the minimisation conditions require $\lambda_4^\mathrm{I}=0$:
\begin{itemize}
\item    C-III-b $(\pm i \hat w_1 ,\, 0 ,\, \hat w_S)$;
\item    C-III-c $(\hat w_1 e^{i \sigma_1},\, \hat w_2 e^{i \sigma_2} ,\, 0)$;
\item    C-III-d $(\pm i \hat w_1  ,\,  \hat w_2 ,\, \hat w_S)$;
\item    C-III-e $(\pm i \hat w_1  ,\, - \hat w_2 ,\, \hat w_S)$;
\item    C-III-f $(\pm i \hat w_1 ,\, i \hat w_2 ,\, \hat w_S)$;
\item    C-III-g $(\pm i \hat w_1 ,\, -i \hat w_2 ,\, \hat w_S)$;
\item    C-III-i $\left(\sqrt{\frac{3(1 + \tan^2 \sigma_1)}{1+9\tan^2 \sigma_1}} \hat w_2 e^{i \sigma_1} ,\, \pm \hat w_2 e^{-i \arctan(3 \tan \sigma_1)} ,\, \hat w_S\right)$;
\item    C-IV-a $\left(\hat w_1 e^{i \sigma_1} ,\, 0 ,\, \hat w_S\right)$;
\item    C-IV-d $\left(\hat w_1 e^{i \sigma_1} ,\, \pm \hat w_2 e^{i \sigma_1} ,\, \hat w_S\right)$;
\item    C-IV-e $\left(\sqrt{-\frac{\sin2 \sigma_2}{\sin2\sigma_1}} \hat w_2 e^{i \sigma_1} ,\, \hat w_2 e^{i \sigma_2} ,\, \hat w_S\right)$;
\end{itemize}

Then, there are models which require $\lambda_4^\mathrm{R}=0$ due to the minimisation conditions, though $\lambda_4^\mathrm{I}$ is left as a free parameter. However, as noted earlier, see Section~\ref{Sec:CP_invariants} (Solution~1), there will be no explicit CP violation present in these models. These models are:
\begin{itemize}
\item \textbf{C-IV-b:}  {\boldmath$\left(\hat w_1 ,\, \pm i \hat w_2 ,\, \hat w_S\right)$}\\
It was pointed out in Ref.~\cite{Emmanuel-Costa:2016vej} that there is no spontaneous CP violation in the exact C-IV-b configuration.

\item  {\boldmath$(\pm \hat w_1e^{-i\arctan(3\tan\sigma_2)}, \sqrt{\frac{3(1+\tan^2\sigma_2)}{1+9\tan^2\sigma_2}}\hat w_1e^{i\sigma_2},\hat w_S)$}\\
Although this vacuum looks like C-III-i, with a change of $w_1 \leftrightarrow w_2$, the origin of it is in C-IV-e, with $\hat w_2 = \sqrt{- \sin(2\sigma_1) / \sin(2\sigma_2)}$ and $\sigma_1 = - \arctan\left(3 \tan \sigma_2\right)$ or \mbox{$\sigma_1 = - \arctan\left(3 \tan \sigma_2\right)+\pi$} when there is a minus sign in the vacuum associated with the first doublet. In the basis with real couplings we get that this vacuum is a special case of C-IV-e. However, when $\lambda_4 \in \mathbb{C}$ the minimisation conditions differ and we can no longer treat this vacuum as a special case of the C-IV-e model. In the C-IV-e like model we have $\lambda_4^\mathrm{I}=0$. 

This model does not violate CP explicitly, however there could be spontaneous CP violation. The C-IV-e model was shown not to violate CP spontaneously due to an overall O(2) rotation of the $S_3$ doublet~\cite{Emmanuel-Costa:2016vej}. Due to $\lambda_4^\mathrm{I} \neq 0$ the O(2) symmetry is no longer present. However, we can perform the basis transformation as given in appendix~\ref{App:BT-complex}.
Both the vacuum and the potential then become real.

\end{itemize}

Finally, we found ``interesting" models when none of the CP-odd invariants, see Section~\ref{Sec:CP_invariants}, vanish:
\begin{itemize}
\item \textbf{C-I-a: \boldmath$(\hat w_1,\,\pm i \hat w_1,\,0)$}\\
There is a single minimisation condition given by
\begin{equation}
\mu_1^2 = -2 \left( \lambda_1 - \lambda_2 \right) \hat w_1^2,
\end{equation}
while $\lambda_4^\mathrm{I}$ is a free parameter.

There are two pairs of mass-degenerate states in the neutral sector.

\item \textbf{C-III-a: \boldmath$(0,\,\hat w_2 e^{i \sigma_2},\, \hat w_S)$}\\
The minimisation conditions are given by
\begin{subequations}
\begin{align}
\mu_{0}^{2}&=-\frac{1}{2} \left( \lambda_5 + \lambda_6 - 2 \lambda_7 \right)\hat{w}_{2}^{2}-\lambda_{8} \hat{w}_{S}^{2} - \lambda_4^\mathrm{I} \frac{\hat w_2^3}{2 \sin\sigma_2 \hat w_S}, \\
\begin{split}\mu_{1}^{2}&=-\left(\lambda_{1}+\lambda_{3}\right) \hat{w}_{2}^{2}-\frac{1}{2}\left[ \lambda_5 + \lambda_6 - \lambda_7 \left( 2 + 8\cos^{2} \sigma_{2}\right) \right] \hat{w}_{S}^{2} - \frac{3}{2} \lambda_4^\mathrm{I} \frac{\hat w_2 \hat w_S}{\sin\sigma_2} , \end{split}\\
\lambda_4^\mathrm{R}&=- \lambda_4^\mathrm{I} \cot \sigma_2 +4\lambda_7\frac{ \cos \sigma_{2} \hat{w}_{S}}{\hat{w}_{2}}.
\end{align}
\end{subequations}

\item \textbf{C-III-h:  \boldmath$(\sqrt{3} \hat w_2 e^{i \sigma_2},\, \pm \hat w_2 e^{i \sigma_2} ,\, \hat w_S)$}\\
The minimisation conditions are given by
\begin{subequations}
\begin{align}
\begin{split}\mu_{0}^{2}&=- 2\left(\lambda_5 + \lambda_6 - 2 \lambda_7\right)\hat{w}_{2}^{2}-\lambda_{8} \hat{w}_{S}^{2}  \pm 4 \lambda_4^\mathrm{I} \frac{\hat w_2^3}{\sin\sigma_2 \hat w_S},\end{split} \\
\begin{split}\mu_{1}^{2}&=-4\left(\lambda_{1}+\lambda_{3}\right) \hat{w}_{2}^{2}-\frac{1}{2}\left[\lambda_5 + \lambda_6 - 2\lambda_{7}\left( 3 + 2 \cos 2\sigma_{2} \right) \right] \hat{w}_{S}^{2} \pm 3 \lambda_4^\mathrm{I} \frac{\hat w_2 \hat w_S}{\sin\sigma_2}, \end{split}\\
\lambda_4^\mathrm{R}&= \mp 2 \lambda_7\frac{\cos\sigma_{2} \hat{w}_{S}}{\hat{w}_{2}}  - \lambda_4^\mathrm{I} \cot\sigma_2 .
\end{align}
\end{subequations}

\item \textbf{C-IV-g: \boldmath$(\hat w_1e^{i\sigma_1},\, \pm i \hat w_1e^{i\sigma_1},\, \hat w_S)$}

This vacuum configuration is not present in the earlier classification of Ref.~\cite{Emmanuel-Costa:2016vej}. When $\lambda_4^\mathrm{I}=0$ this vacuum reduces to either C-III-i with $\sigma_1 = \pm \pi/6$ or C-IV-e. In both cases, C-III-i and C-IV-e, there is no spontaneous CP violation.

When we allow for a complex $\lambda_4$, the minimisation conditions are given by
\begin{subequations}
\begin{align}
\mu_0^2 &= - (\lambda_5+\lambda_6)\hat w_1^2-\lambda _8 \hat w_S^2,\\
\mu_1^2&=  -2\left( \lambda _1- \lambda _2\right)\hat w_1^2-\frac{1}{2} (\lambda_5 + \lambda_6) \hat w_S^2, \\
\lambda_4^\mathrm{R} &=\pm \frac{\sin3\sigma_1 \hat w_S}{\hat w_1}\lambda_7,\\ 
\lambda_4^\mathrm{I} &=\pm \frac{\cos3\sigma_1 \hat w_S}{\hat w_1}\lambda_7.
\end{align}
\end{subequations}

By analysing the scalar sector we found that the neutral mass-squared matrix always results in at least two negative eigenvalues. These are not caused by the perturbativity limit of the scalar quartic couplings. Introduction of soft symmetry breaking terms in the scalar potential would solve the issue. Another possibility would be to set $\lambda_7=0$. However, in that case one of the scalars becomes massless.

\item \textbf{C-V: \boldmath$\left(\hat w_1 e^{i \sigma_1} ,\, \hat w_2 e^{i \sigma_2} ,\, \hat w_S\right)$}\\
The minimisation conditions are given by
\begin{subequations}\label{Eq:CV_min_cond}
\begin{align}
\begin{split}
\mu_{0}^{2}&= - \frac{1}{2} \left( \lambda_5 + \lambda_6\right) \left( \hat{w}_1^2 + \hat{w}_2^2 \right)+ \lambda_4^\mathrm{I} \frac{\hat{w}_2}{\hat{w}_S} \frac{C_1}{C_2} - \lambda_8 \hat{w}_S^2
,\end{split}\\
\mu_{1}^{2}&=-\left(\lambda_{1}-\lambda_{2}\right)\left(\hat{w}_{1}^{2}+\hat{w}_{2}^{2}\right) - \lambda_4^\mathrm{I}\frac{\hat{w}_S}{\hat{w}_2} \frac{C_3}{C_4}-\frac{1}{2}\left(\lambda_{5}+\lambda_{6}\right) \hat{w}_{S}^{2},\\
\lambda_{2}+\lambda_{3}&= \lambda_4^\mathrm{I} \frac{\hat{w}_S}{\hat{w}_2} \frac{C_5}{C_6},\\
\lambda_4^\mathrm{R}&= \lambda_4^\mathrm{I} \frac{ \sin \sigma_1 \hat{w}_1^2 - \left[ 2 \sin \sigma_1 - \sin\left( \sigma_1 - 2 \sigma_2 \right) \right] \hat{w}_2^2 }{\cos \sigma_1 \hat{w}_1^2 - \left[ 2 \cos \sigma_1 + \cos \left( \sigma_1 - 2 \sigma_2 \right) \right] \hat{w}_2^2},\\
\lambda_7&= - \lambda_4^\mathrm{I} \frac{\hat{w}_2}{\hat{w}_S} \frac{C_7}{C_2},
\end{align}
\end{subequations}
with
\begin{subequations}\label{Eq:CV_min_cond-Cs}
\begin{align}
\begin{split}
C_1 &= \cos\left( \sigma_1 - 3 \sigma_2 \right) \hat{w}_2^6\\
&\quad - \left[ \cos\left( 3 \sigma_1 - 5 \sigma_2 \right)  + 8 \cos\left( \sigma_1 - \sigma_2 \right) \cos 2 \sigma_2 \right] \hat{w}_1^2 \hat{w}_2^4\\
&\quad+ \cos\left( 3 \sigma_1 - \sigma_2 \right) \hat{w}_1^2 \left( 2 \hat{w}_1^4 - 5 \hat{w}_1^2 \hat{w}_2^2 + 2 \hat{w}_2^4 \right)\\
&\quad + \cos\left( \sigma_1 + \sigma_2 \right) \left( \hat{w}_1^6 - 2 \hat{w}_1^4 \hat{w}_2^2 + 2 \hat{w}_2^6 \right),
\end{split}\\
\begin{split}
C_2 &= 2 \left\lbrace \cos \sigma_1 \hat{w}_1^2  - \left[2 \cos \sigma_1 + \cos \left( \sigma_1 - 2 \sigma_2 \right)\right]  \hat{w}_2^2 \right\rbrace\\
&\quad\times \left\lbrace \sin 2 \sigma_1  \hat{w}_1^2 + \sin 2 \sigma_2  \hat{w}_2^2 \right\rbrace,
\end{split}\\
C_3&= \left\lbrace  \hat w_1^2 - \left[ 2 + \cos\left( 2 \sigma_1 - 2 \sigma_2 \right) \right] \hat w_2^2 \right\rbrace^2 + \sin^2\left( 2\sigma_1 - 2 \sigma_2\right) \hat w_2^4,\\
C_4&= 4 \sin\left( \sigma_1 - \sigma_2 \right) \left\lbrace \cos \sigma_1 \hat{w}_1^2 - \left[ 2 \cos \sigma_1 + \cos \left( \sigma_1 - 2 \sigma_2 \right) \right]\hat{w}_2^2 \right\rbrace,\\
\begin{split}
C_5 &= 2 \cos  2 \sigma_1  \sin \left( 2 \sigma_1 - 2 \sigma_2 \right) \hat{w}_1^2 \hat{w}_2^2 + 2 \left[ \sin 2 \sigma_2  - \sin \left( 2 \sigma_1 - 4 \sigma_4 \right) \right] \hat{w}_2^4\\
&\quad + \sin  2 \sigma_1  \left( \hat{w}_1^4 - 6 \hat{w}_1^2 \hat{w}_2^2 + 5 \hat{w}_2^4 \right),
\end{split}\\
C_6 &= C_4 \left[ \sin 2 \sigma_1  \hat{w}_1^2 + \sin 2 \sigma_2  \hat{w}_2^2 \right],\\
C_7 &= \cos\left( \sigma_1 - \sigma_2 \right) \left\lbrace 3 \hat{w}_1^4 - \left[ 8 + 2 \cos \left( 2 \sigma_1 - 2 \sigma_2 \right) \right] \hat{w}_1^2 \hat{w}_2^2 + 3 \hat{w}_2^4 \right\rbrace.
\end{align}
\end{subequations}
While the dependence of the potential on the phases of the vevs is rather complicated, we should like to stress two facts:
\begin{itemize}
\item
When both $\sigma_1$ and $\sigma_2$ approach zero (the limit of real vevs), several things happen: (1) $\lambda_4^\mathrm{R}\to0$, (2) $C_2\to0$, and (3) $C_6\to0$. The two latter points lead to $(\lambda_2+\lambda_3)$ and $\lambda_7$ both diverging and exceeding the perturbativity limit. In the truly real case this is avoided by having $\lambda_4^\mathrm{I}=0$.
\item
$C_2$ and $C_6$ both contain a factor that vanishes for $\sin2\sigma_2=\sin2\sigma_1(\hat w_1^2/\hat w_2^2)$, a factor that is periodic in $\sigma_2$, with period $\pi$.
\end{itemize}

In the case of a real scalar potential there were three massless neutral states present due to the $O(2) \otimes U(1)_{h_1} \otimes U(1)_{h_2} \otimes U(1)_{h_S}$ symmetry~\cite{Kuncinas:2020wrn}. In a basis with complex couplings this is no longer so. When $\lambda_4^\mathrm{I} = 0$ the minimisation conditions require $\lambda_{23} = \lambda_4 = \lambda_7 = 0$. This allows to rotate away all of the phases and the model becomes CP-conserving~\cite{Emmanuel-Costa:2016vej}.

\end{itemize}

\section{Yukawa Lagrangian}\label{Sec:Yukawa_Lagrangian}

The $S_3$ symmetry within the fermionic sector in the context of 3HDMs was previously studied~\cite{Kubo:2003iw,Teshima:2005bk,Mondragon:2007nk,Mondragon:2007af,Teshima:2012cg,GonzalezCanales:2012blg,Ma:2013zca,GonzalezCanales:2013pdx,Das:2015sca,Cruz:2017add}. There are several possibilities to group fermions into the $S_3$ tuplets. Whenever the singlet vev, $w_S$, is different from zero we can group fermions into a trivial singlet representation. This case yields the fermion mass matrices:
\begin{subequations} \label{Eq.FMMws}
\begin{align}
\mathcal{M}_u={}&\frac{1}{\sqrt{2}} \left( y_{ij}^u \right) w^*_S,\\
\mathcal{M}_d={}&\frac{1}{\sqrt{2}} \left(y_{ij}^d \right )w_S,
\end{align}
\end{subequations}
where the $y$s are the Yukawa couplings, which can be complex, and are not constrained by the $S_3$ symmetry.

Another possibility relies on assigning non-trivial $S_3$ charges to fermions. It is possible to group fermions into a singlet-doublet representation, as in the case of the scalar SU(2) doublets,
\begin{equation*}
\mathbf{2}:\left(Q_1\,Q_2\right)^\mathrm{T},\,\left(u_{1R}\,u_{2R}\right)^\mathrm{T},\,\left(d_{1R}\,d_{2R}\right)^\mathrm{T}\quad\text{and}\quad\mathbf{1}:Q_3,\,u_{3R},\,d_{3R}.
\end{equation*}
We have freedom to assign generations of fermions to different $S_3$ representations, and not in a strictly increasing order, e.g., $Q_1$ might be associated with the third fermionic family, rather than the first one. Such structure yields the mass matrix for each quark sector ($d$ and $u$) of the form
\begin{subequations} \label{Eq:Mf}
\begin{align}
{\cal M}_u=
\frac{1}{\sqrt{2}}\begin{pmatrix}
y_1^uw_S^\ast+y_2^u w_2^\ast & y_2^u w_1^\ast & y_4^u w_1^\ast \\
y_2^u w_1^\ast & y_1^u w_S^\ast-y_2^u w_2^\ast & y_4^u w_2^\ast \\
y_5^u w_1^\ast & y_5^u w_2^\ast & y_3^u w_S^\ast
\end{pmatrix},\\
{\cal M}_d=
\frac{1}{\sqrt{2}}\begin{pmatrix}
y_1^dw_S+y_2^dw_2 & y_2^d w_1 & y_4^d w_1 \\
y_2^d w_1 & y_1^d w_S-y_2^d w_2 & y_4^d w_2 \\
y_5^d w_1 & y_5^d w_2 & y_3^d w_S
\end{pmatrix}\label{Eq:SS_rep_D}.
\end{align}
\end{subequations}
In accordance with complex couplings present in the scalar potential, we shall assume that the Yukawa couplings could, in principle, also be complex.

Apart from the singlet representation there is a pseudo-singlet representation in $S_3$. Although we assume that the scalar potential is given by a singlet-doublet representation this does not force the Yukawa sector to be presented strictly in terms of singlets. There is the possibility to group some of the fermions (the ones with subindices ``3") into the $S_3$ pseudo-singlet representation; while he $S_3$ doublet representation stays intact. In this case there are three additional possibilities to construct the fermionic mass matrices. For simplicity we shall present only the $\mathcal{M}_d$ mass matrix,
\begin{subequations}\label{Eq:Other_rep_d}
\begin{align}
\mathbf{1^\prime}:Q_3,\,d_{3R}:\qquad  &\mathcal{M}_d=
\frac{1}{\sqrt{2}}\begin{pmatrix}
y_1^dw_S+y_2^dw_2 & y_2^d w_1 & y_4^d w_2 \\
y_2^d w_1 & y_1^d w_S-y_2^d w_2 & -y_4^d w_1 \\
y_5^d w_2 & -y_5^d w_1 & y_3^d w_S
\end{pmatrix},\\
\mathbf{1^\prime}:Q_3,~\mathbf{1}:\,d_{3R}:\qquad  &\mathcal{M}_d=
\frac{1}{\sqrt{2}}\begin{pmatrix}
y_1^dw_S+y_2^dw_2 & y_2^d w_1 & y_4^d w_1 \\
y_2^d w_1 & y_1^d w_S-y_2^d w_2 & y_4^d w_2 \\
y_5^d w_2 & -y_5^d w_1 & 0
\end{pmatrix},\label{Eq:MD_ps_s}\\
\mathbf{1}:Q_3,~\mathbf{1^\prime}:\,d_{3R}:\qquad  &\mathcal{M}_d=
\frac{1}{\sqrt{2}}\begin{pmatrix}
y_1^dw_S+y_2^dw_2 & y_2^d w_1 & y_4^d w_2 \\
y_2^d w_1 & y_1^d w_S-y_2^d w_2 & -y_4^d w_1 \\
y_5^d w_1 & y_5^d w_2 & 0
\end{pmatrix}.\label{Eq:MD_s_ps}
\end{align}
\end{subequations}
The mixed representations, eq.~\eqref{Eq:MD_ps_s} and eq.~\eqref{Eq:MD_s_ps}, are equivalent. Without loss of generality we can change the overall sign of the Yukawa couplings. Let us consider $\mathcal{M}_d$ of eq.~\eqref{Eq:MD_s_ps} with $\{y_2^d,\, y_5^d\} \to - \{y_2^d,\, y_5^d\}$. Next, we can rotate the mass matrix by
\begin{equation}
U =  \begin{pmatrix}
0 & -1 & 0 \\
1 & 0 & 0 \\
0 & 0 & 1
\end{pmatrix}.
\end{equation}
This way we get from $\mathcal{M}_d$ of eq.~\eqref{Eq:MD_s_ps} to $\mathcal{M}_d$ of eq.~\eqref{Eq:MD_ps_s}. An interesting aspect to note of the mixed representations is that the number of Yukawa couplings is reduced from five to four, $y_3^d=0$.

Apart from assigning different chirality states to different representations,
\begin{equation*}
\mathbf{1}:Q_3,~\mathbf{1^\prime}:\,u_{3R},\,\,d_{3R}, \qquad 
\mathbf{1^\prime}:Q_3,~\mathbf{1}:\,u_{3R},\,\,d_{3R},
\end{equation*}
we could also assign the up- and down-quarks to different representations: \begin{equation*}
\begin{aligned}[c]
\mathbf{1}:Q_3,~\mathbf{1}:\,u_{3R},~\mathbf{1^\prime}:\,d_{3R},\\
\mathbf{1}:Q_3,~\mathbf{1^\prime}:\,u_{3R},~\mathbf{1}:\,d_{3R},
\end{aligned}
\qquad
\begin{aligned}[c]
\mathbf{1^\prime}:Q_3,~\mathbf{1}:\,u_{3R},~\mathbf{1^\prime}:\,d_{3R},\\
\mathbf{1^\prime}:Q_3,~\mathbf{1^\prime}:\,u_{3R},~\mathbf{1}:\,d_{3R}.
\end{aligned}
\end{equation*}
In total, there are eight different possibilities to construct the Yukawa sector by assigning different $S_3$ charges to fermions: four different possibilities in the down sector, eqs.~(\ref{Eq:SS_rep_D}), (\ref{Eq:Other_rep_d}) times two different possibilities for the up sector, since the $S_3$ charge of $Q_3$ is fixed by the down sector. However, the number of independent solutions is reduced by one as two representations yield identical results, $(\mathbf{1}:Q_3,~\mathbf{1^\prime}:\,u_{3R},\,\,d_{3R}) \text{ and } (\mathbf{1^\prime}:Q_3,~\mathbf{1}:\,u_{3R},\,\,d_{3R}).$

After defining the fermionic sector we can proceed to check if some of the models could generate realistic fermionic masses. The cases with explicit or spontaneous CP violation are:

\begin{itemize}
\item \textbf{R-I-1}

As the vacuum is given by $\{0,\,0,\,w_S\}$, the only possibility is to group fermions into the $S_3$ singlets. The pseudo-singlet representation would be equivalent to the representation which is given in terms of singlets. The complex CKM parameters would then be generated by complex Yukawa couplings.

\item \textbf{R-I-2a} and \textbf{R-I-2b,c}

When fermions are grouped into identical singlet or pseudo-singlet representations it is not possible to generate realistic fermionic masses as $\det\left(\mathcal{M}_f\right)=0$, which indicates that one of the fermions is massless. However, when fermions are grouped into mixed representations of singlets and pseudo-singlets, $\det\left(\mathcal{M}_f\right) \neq 0$. Nevertheless, due to the form of the eigenvalues it is impossible to fit realistic masses as those imply that two of the fermions would have nearly identical masses.

\item \textbf{C-I-a}

Regardless of the $S_3$ charges of the fermions the Hermitian mass-squared matrix is always given by
\begin{equation}
\mathcal{H}_d = \mathcal{M}_d \mathcal{M}_d^\dagger = \frac{\hat w_1^2}{2}\begin{pmatrix}
|y_4^d|^2 + 2 |y_2^d|^2 & \pm i \left( 2|y_2^d|^2 -|y_4^d|^2 \right) & 0 \\
\mp i \left( 2|y_2^d|^2 -|y_4^d|^2 \right) &  |y_4^d|^2 + 2 |y_2^d|^2 & 0 \\
0 & 0 & 2|y_5^d|^2
\end{pmatrix}.
\end{equation}
From this form of the mass-squared matrix it is obvious that the CKM matrix is unrealistic.

\item \textbf{C-III-a}

After a unitary transformation (only in the case of the pseudo-singlet representation) it is possible to write the mass matrix for both singlet and pseudo-singlet representations as
\begin{equation}
\mathcal{M}_d = \frac{1}{\sqrt{2}}\begin{pmatrix}
e^{i \sigma_2} y_2^d \hat w_2 + y_1^d \hat w_S & 0 & 0 \\
0 & -e^{i \sigma_2} y_2^d \hat w_2 +  y_1^d \hat w_S &  e^{i \sigma_2} y_4^d \hat w_2 \\
0 & e^{i \sigma_2} y_5^d \hat w_2 &  y_3^d \hat w_S
\end{pmatrix},
\end{equation}
Due to the block-diagonal form of the matrix it is impossible to generate a realistic CKM matrix. The mixed representations yields $\det\left(\mathcal{M}_f\right)=0$. 

Another possibility would be to construct a trivial Yukawa sector. As in the case of R-I-1 the Yukawa couplings would need to be complex.

\item \textbf{C-III-h}

When the singlet representation is considered, the mass matrix can be rotated by
\begin{equation}
U = \begin{pmatrix}
\cos \theta & \sin \theta & 0 \\
-\sin \theta & \cos \theta & 0 \\
0 & 0 & 1
\end{pmatrix},
\end{equation}
with $\theta = - \pi/3$, so that 
\begin{equation}
U \mathcal{M}_d U^\mathrm{T} = \frac{1}{\sqrt{2}} \begin{pmatrix}
- 2 e^{i \sigma_2} y_2^d \hat w_2 + y_1^d \hat w_S & 0 & 0 \\
0 & 2 e^{i \sigma_2} y_2^d \hat w_2 + y_1^d \hat w_S & 2 e^{i \sigma_2} y_4^d \hat w_2 \\
0 & 2 e^{i \sigma_2} y_5^d \hat w_2 & y_3^d \hat w_S
\end{pmatrix}.
\end{equation}
In the case of the pseudo-singlet representation it is possible to get an identical matrix by choosing $\theta = - 5\pi/6$ and changing the sign $y_2^d \to -y_2^d$. In these cases it is not possible to generate a realistic CKM matrix.

The mixed representation results in $\det\left(\mathcal{M}_f\right)=0$. The only viable possibility is to consider a trivial Yukawa sector.

\end{itemize}

\section{Numerical studies}\label{Sec:Num_exploration}
The remaining cases (C-IV-c, C-IV-f, C-IV-g, C-V) require further discussion as those were explored numerically. A complete systematic study of the models presented here, involving both the scalar and Yukawa sectors, is beyond the scope of this work. We rely on a simplified check of the models at tree level (and also disregarding the leptonic sector) to draw a conclusion if a specific model can be rejected or not. In total, we fit several parameters, adopting a 3-$\sigma$ tolerance of values taken from the PDG~\cite{ParticleDataGroup:2022pth}:
\begin{itemize}
\item Masses of the up- and down-quarks;
\item The absolute values, arguments of the unitarity triangle ($\alpha,\,\sin 2\beta,\,\gamma$) and independent measure of CP violation ($J$)~\cite{Jarlskog:1985ht,Bernabeu:1986fc} of the CKM matrix;
\end{itemize}
For the C-V model, we perform several additional checks. These are:
\begin{itemize}
\item Interactions of the SM-like Higgs boson with fermions. We assume the Higgs boson signal strength in the $b$-quark channel~\cite{CMS:2020zge,ATLAS:2020fcp,ATLAS:2020jwz} as a reference point and apply the corresponding limits to other channels;
\item Suppressed scalar mediated FCNC~\cite{CMS:2021gfa,ATLAS:2022gzn};
\item CP properties of the SM-like Higgs boson~\cite{ATLAS:2020ior,CMS:2022dbt};
\item Upper limit on the decay of the $t$-quark into lighter charged scalars when decays are not kinematically suppressed~\cite{CMS:2020osd,ATLAS:2021zyv};
\end{itemize}
The additional checks are performed only in case of C-V as those demand input from the scalar sector; to be more precise, one needs to know how the scalar mass eigenstates look like to determine the interaction strength of the SM-like Higgs boson with the fermions. Neither of the other models (C-IV-c, C-IV-f, C-IV-g) can generate a realistic scalar content, unless the $S_3$ symmetry is softly broken. We do not consider soft symmetry breaking.

Fitting the discussed constraints is performed by taking vevs as inputs and scanning over ten free Yukawa couplings ($y_i^u,\,y_i^d$) in the range $\{|y_i^u|,\,|y_i^d|\} \leq \sqrt{4\pi}$. Only one (one for the up-quarks and one for the down-quarks) of the Yukawa couplings (largest in absolute value) tends to be in the range of $0.1 \leq\{|y_i^u|,\,|y_i^d|\} \leq 1.5$, while others are within $\mathcal{O}(10^{-10}) - \mathcal{O}(10^{-3})$. It is also possible to achieve a valid (simplified) fit, when two out of the ten Yukawa couplings vanish. However, in this case there will be less freedom to fit the scalar-fermion interactions.

The strategy adopted for the fitting was as follows.
Due to a non-linear fitting function we found that the easiest approach is to optimise the numerical values of the Yukawa couplings by utilising the gradient descent and random search techniques. Since the optimisation of the scalar-fermionic sector is a computationally expensive task time-wise, a simplified scan was performed. This involves performing a scan over the constraints by binning the free variables of vevs ($n$) into approximately $10^3$ $n$-dimensional boxes. Then, the vevs are kept fixed while only the values of ($y_i^u,\,y_i^d$) are evolved. If the fitting procedure would take longer than a set threshold (one hour) a new vev would be chosen within the same bin. Also, the scan was performed over all possible combinations of the up- and down-quarks, left- and right-chiral states having different $S_3$ charges assigned. Scanning over all representations might seem to be a redundant procedure since some representations are more constraining (can have vanishing Yukawa couplings) than others. However, this could also be viewed as a wider coverage of the available parameter space to balance the poor $10^3$ data sampling of vevs. Regardless, we would like to emphasize that the reader should be cautious when interpreting the results (especially the C-IV-c, C-IV-f and C-IV-g models) since the checks were performed using a rather scarce grid and not all of the experimental data were fitted. 

We shall next comment on models which could generate a non-trivial Yukawa sector:
\begin{itemize}

\item \textbf{C-IV-c} 

In the case of real Yukawa couplings it is possible to fit both the fermionic masses and the CKM matrix in nearly all of the $S_3$ representations. When the $(\mathbf{1}:Q_3,~\mathbf{1^\prime}:\,u_{3R},\,\,d_{3R})$ representation, or equivalently  $(\mathbf{1^\prime}:Q_3,~\mathbf{1}:\,u_{3R},\,\,d_{3R})$, is considered it is not possible to achieve a 3-$\sigma$ fit for all of the CKM considered experimental constraints. With complex Yukawa couplings (assuming at least nine more degrees of freedom) this is no longer true. However, C-IV-c requires $\lambda_4^\mathrm{I}=0$ and one might consider it ``more natural" to take real Yukawa couplings in this case.

Nevertheless, there exists an accidental (not the Goldstone boson due to a broken continuous symmetry) massless scalar state in the model, see Ref.~\cite{Kuncinas:2020wrn}. Soft symmetry breaking could be introduced to make it massive.

\item \textbf{C-IV-f}

The C-IV-c model is contained within C-IV-f with $\sigma_1=0$. Therefore, the results do not differ significantly. There shall be more freedom due to an additional phase coming from vevs, $\sigma_1$.

A massless state is also present in the C-IV-f model, and hence a possible solution is to introduce soft symmetry breaking.

\item \textbf{C-IV-g}

The results of fitting the Yukawa sector are identical to C-IV-c since both vevs are described in terms of three parameters.

This model is only possible when $\lambda_4^\mathrm{I} \neq 0$. There are negative mass-squared scalars present. Introduction of soft symmetry breaking terms could possibly produce non-negative eigenvalues.

\item \textbf{ C-V }

This is the only model out of those studied that could yield a realistic scalar sector without the need to introduce soft breaking terms. In light of that, several tree-level constraints were imposed on the scalar sector at the 3-$\sigma$ level:
\begin{itemize}
\item Perturbativity (quartic scalar-scalar couplings), stability and unitarity. The unitarity constraints are discussed in Appendix~\ref{App:Unitarity};
\item LEP constraints. We adopt a generous lower bound for charged scalars, $m_{\varphi_i^\pm} \geq 70\text{ GeV}$~\cite{Pierce:2007ut,Arbey:2017gmh}. Decays of the $W^\pm$ and $Z$ into a pair of scalars are kinematically suppressed~\cite{Schael:2013ita};
\item Mass of the SM-like Higgs boson $h$;
\item Decay of the SM-like Higgs boson into lighter scalars is assumed to be within $\mathrm{Br(h\to \text{BSM}) \lesssim 0.1}$. This constraint is approximated by fixing the total width of the SM-like Higgs boson to be $\Gamma_h^\mathrm{tot} = 6\,\text{MeV}$;
\item SM-like scalar-gauge couplings;
\item Electroweak precision observables~\cite{Peskin:1990zt,Peskin:1991sw} using techniques of Refs.~\cite{Grimus:2007if,Grimus:2008nb};
\end{itemize}
In the above checks we assume that the SM-like Higgs boson-fermion interactions are close to the SM-like values. This is done due to many degrees of freedom coming from the fermionic interactions. The Yukawa sector is evaluated separately, and shall be discussed below. Due to such limitations we do not discuss constraints relying on the scalar-fermions interactions, like the di-photon partial width, or $B$-physics or the electric dipole moment. Also, the SM-like Higgs decay into additional scalars is only roughly approximated.

Vacuum expectation values can be parameterised in terms of two angles:
\begin{equation}
\hat w_1 = v \sin \alpha \cos \beta, \quad \hat w_2 = v \cos \alpha \cos \beta, \quad \hat w_S = v \sin \beta.
\end{equation}
The available parameter space of the vevs after applying the discussed constraints is shown in Fig.~\ref{Fig:C-V_VEVs}. We note that the central point (left panel), $\alpha=\beta=\pi/4$ corresponds to $\hat w_1^2=\hat w_2^2=v^2/4$, with $\hat w_S^2=v^2/2$. But there is a considerable spread around this point. A striking feature of the right panel is the fact that there are alternating bands of allowed and excluded region where $\sigma_2\simeq\sigma_1\pm n\pi$. This can be understood in terms of the discussion following eq.~(\ref{Eq:CV_min_cond-Cs}).

\begin{figure}[htb]
\begin{center}
\includegraphics[scale=0.275]{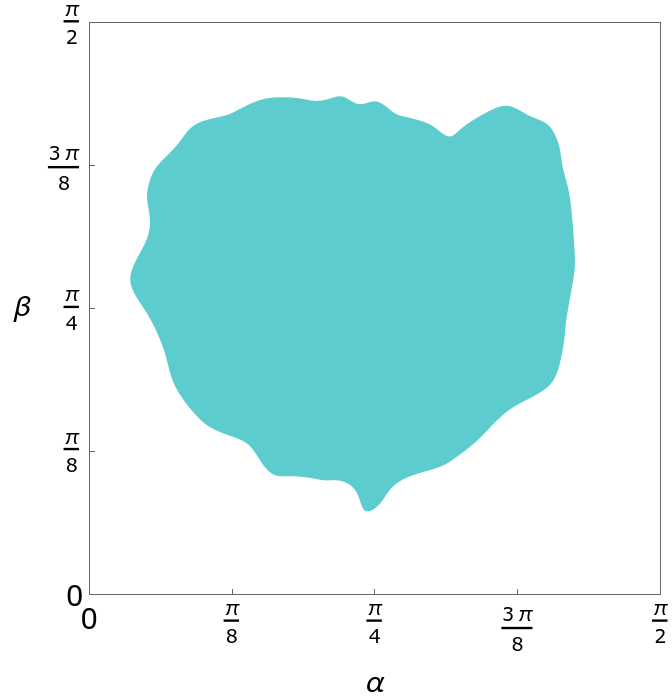}\hspace*{10pt}
\includegraphics[scale=0.275]{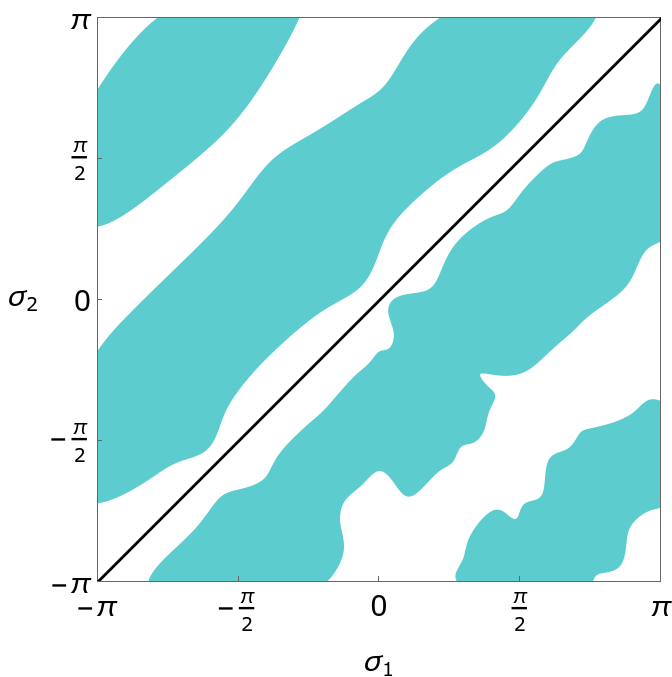}
\end{center}
\caption{Scatter plots of vevs in the C-V model after imposing constraints. Left: the absolute values of vevs parameterised in terms of two angles. Right: phases of the vevs.}
\label{Fig:C-V_VEVs}
\end{figure}

In the model there are two charged scalars, $H_i^\pm$. The neutral sector consists of five scalars ($h,H_1, H_2, H_3, H_4$) which all mix. This results in CP-indefinite states. The scalars $H_i$ are ordered from the lightest to the heaviest states, with masses $m_{H_i} \leq m_{H_{i+1}}$. The $h$ state is required to be compatible with the SM-like Higgs boson. We assume that $h$ can be in any position of sorted masses as long as it is close to 125 GeV, e.g., the 125 GeV state could be lighter than $H_1$ or heavier than $H_4$. The mass-scatter plots are presented in Fig.~\ref{Fig:C-V_masses}. The dashed lines indicate possible locations of the SM-like $h$, for the different hierarchies.

\begin{figure}[htb]
\begin{center}
\includegraphics[scale=0.275]{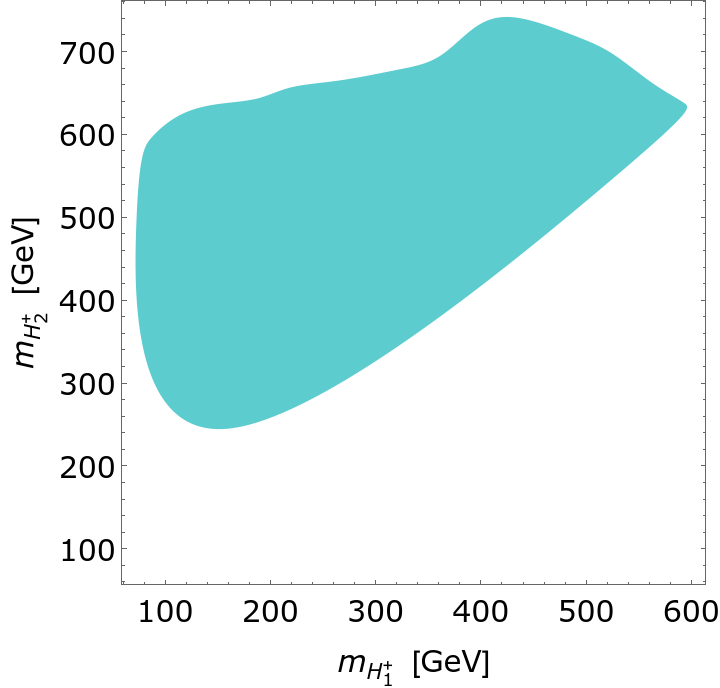}
\includegraphics[scale=0.275]{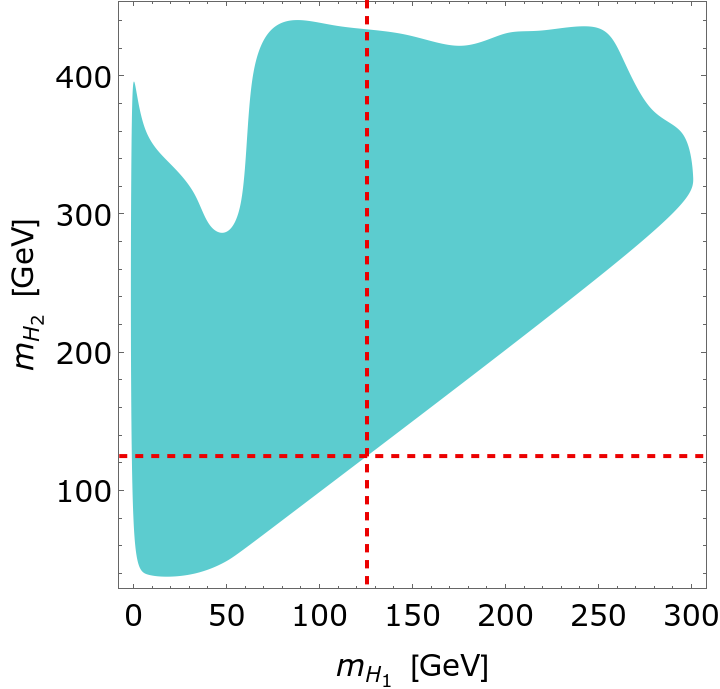}
\includegraphics[scale=0.275]{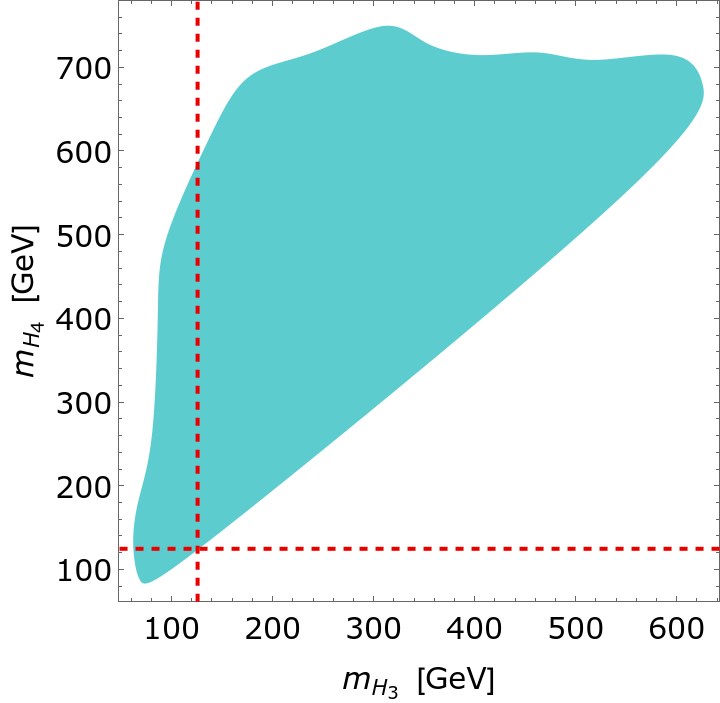}
\end{center}
\caption{Scatter plots of masses that satisfy constraints in the C-V model. Left: the charged sector, $H_i^\pm$. Middle and right: the active sector, $H_i$. In the neutral sector the red line indicates a 125 GeV state. Possible locations of the SM-like $h$ are indicated by the dashed lines.}
\label{Fig:C-V_masses}
\end{figure}

The alignment of masses $m_{H_i}$ suggests that the SM-like Higgs boson could be the heaviest one out of the neutral states. We acknowledge the fact that some parameter space of the model is left unconstrained as not all of the LHC-related constraints are considered. Nevertheless, it is of interest to consider light scalars~\cite{Fox:2017uwr,Cepeda:2021rql,Biekotter:2022jyr,Robens:2022zgk,Plantey:2022jdg}, which are present in many extensions of the scalar potential.

Quite a remarkable aspect of the model is the presence of $m_{H_1} = \mathcal{O}(\text{MeV})$ neutral scalar states, which are not excluded by the constraints. Due to freedom of the model it is possible to have suppressed decays of the SM-like $h$ state into the $H_1$ states, $g(h H_1 H_1) \sim \mathcal{O}(10^{-6})$. We shall focus our discussion on the sub-GeV states. The light states arise when $|\lambda_4| \lesssim \mathcal{O}(10^{-3})$ and $\lambda_7 \lesssim \mathcal{O}(10^{-5})$. A naive idea could be that this case becomes comparable to the one (C-V) with $\lambda_4^\mathrm{I}=0$. However, this is not true since $0.1 < \lambda_{23} < 3$ in the discussed parameter space and the minimisation condition of the real scalar potential would require $\lambda_{23}=0$. Note that the discussed light states are specific to C-V with $\lambda_4$ complex, in C-V with real $\lambda_4$ there are three unwanted Goldstone states present.

With input from the scalar sector one can check if the constraints coming from the Yukawa Lagrangian are satisfied. Since the fermionic masses and the CKM-related fits can be satisfied for more constrained models (C-IV-c, C-IV-f, C-IV-g), of primary interest would be the scalar-fermion couplings. As in the case of other models, we performed a limited scan, however this time the available parameter space of vevs was constrained. With real Yukawa couplings it is possible to fit any representation while satisfying the additional scalar-fermion constraints.

Interactions of the neutral non-SM scalars $H_i$ with fermions are left unconstrained. Unlike in the SM case, where $g(h f_i \bar f_i) = - i\,m_{f_i}/v$, due to mixing the $g(H_i f_j \bar f_k)$ couplings are not proportional to masses of the fermions. It becomes possible to generate scalar-fermion interactions with ``diagonal'' decays to the same species of fermions suppressed over FCNC, $g(H_i f_j \bar f_k) \gg g(H_i f_j \bar f_j)$, or several interactions might be prohibited in part of the parameter space, $g(H_i f_j \bar f_k)=0$, assuming also $j=k$. This is also true for the $\mathcal{O}(\text{MeV})$ states. It might be possible to have really low fermionic decay widths so that the light scalars would escape the detector. The scalar-fermion couplings should be constrained when loop-induced observables are considered, $e.g.$, di-photon signals, $B$-physics, electric dipole moment, $(g-2)_\mu$. One of the possibilities could be to consider complex Yukawa couplings, which, depending on the representation, could introduce from eight to ten additional degrees of freedom.

\end{itemize}

Results of different $S_3$ charges assigned to fermions are summarised in Table~\ref{Table:CPV_models_list}. A large number of cases of Ref.~\cite{Emmanuel-Costa:2016vej} in our classification was rejected due to the minimisation conditions forcing vanishing of new CP sources, to be more precise the $\lambda_4^\mathrm{I}=0$ condition. Although we do not consider soft symmetry breaking in the scalar potential, it might be interesting to consider different $S_3$ structures of the Yukawa Lagrangian with such vacua. 

{{\renewcommand{\arraystretch}{1.15}
\begin{table}[htb]
\caption{A summary of different CP violating models. In the first column we list if the scalar potential can be complex. The CPV column indicates whether there is spontaneous CP violation, as presented in Ref.~\cite{Emmanuel-Costa:2016vej} with $\lambda_4^\mathrm{I}=0$, or explicit, which requires strictly complex $\lambda$s. In the last column possible structures of the Yukawa Lagrangian ($\mathcal{L}_Y$) are presented. Entries with ``-" indicate that it is not possible to construct a realistic $\mathcal{L}_Y$.}
\label{Table:CPV_models_list}
\begin{center}
\begin{threeparttable}
\begin{tabular}{|c|c|c|c|c|}\hline\hline
\begin{tabular}[c]{@{}c@{}} Scalar\\ potential \end{tabular} & Vacuum & vevs & CPV & $\mathcal{L}_Y$ \\ \hline\hline
complex & R-I-1 & $(0,\,0,\,w_S)$ & explicit & trivial \\ \hline
complex & R-I-2a & $(w_1,\,0,\,0)$ & explicit &  - \\ \hline
complex & R-I-2b,c & $(w_1,\, \pm \sqrt{3} w_1,\,0)$ & explicit & - \\ \hline\hline
complex & C-I-a & $(\hat w_1,\,\pm i \hat w_1,\,0)$ & explicit &  - \\ \hline
\begin{tabular}[l]{@{}c@{}} complex \\ real \end{tabular} & C-III-a & $(0,\,\hat w_2 e^{i \sigma_2},\, \hat w_S)$ &\begin{tabular}[c]{@{}c@{}} explicit\\ spontaneous \end{tabular} & trivial \\ \hline
\begin{tabular}[l]{@{}c@{}} complex\\ real \end{tabular} & C-III-h & $(\sqrt{3} \hat w_2 e^{i \sigma_2},\, \pm \hat w_2 e^{i \sigma_2} ,\, \hat w_S)$ &\begin{tabular}[c]{@{}c@{}} explicit\\ spontaneous \end{tabular} & trivial \\ \hline
real$^\alpha$ & C-IV-c & $\left(\sqrt{1 + 2 \cos^2 \sigma_2} \hat w_2 ,\, \hat w_2 e^{i \sigma_2} ,\, \hat w_S\right)$ & spontaneous & any \\ \hline
real$^\alpha$ & C-IV-f & $\left(\sqrt{2 + \frac{\cos(\sigma_1 - 2\sigma_2)}{\cos \sigma_1}} \hat w_2 e^{i \sigma_1},\, \hat w_2 e^{i \sigma_2} ,\, \hat w_S\right)$ &  spontaneous & any \\ \hline
complex$^\beta$ & C-IV-g & $(\hat w_1e^{i\sigma_1},\, \pm i \hat w_1e^{i\sigma_1},\, \hat w_S)$ & explicit  & any \\ \hline
complex & C-V & $\left(\hat w_1 e^{i \sigma_1} ,\, \hat w_2 e^{i \sigma_2} ,\, \hat w_S\right)$ & explicit & any \\ \hline
\end{tabular}
  \footnotesize
  \begin{tablenotes}
  \item[$\alpha$] In C-IV-c and C-IV-f there is a massless scalar present. Soft symmetry breaking would remove the massless scalar.
  \item[$\beta$] C-IV-g results in at least two negative mass-squared eigenvalues. Introduction of soft symmetry breaking terms might solve the issue.
  \end{tablenotes}
  \end{threeparttable}
\end{center}
\end{table}}

The soft terms are as follows:
\begin{equation}\label{Eq:VSoftGenericBasis}
\begin{aligned}
V_2^\prime ={}&\mu_2^2 \left( h_1^\dagger h_1 - h_2^\dagger h_2 \right) 
+ \frac{1}{2}  \left(\nu_{12}^2\, h_1^\dagger h_2 + \mathrm{h.c.} \right)\\&
+ \frac{1}{2}  \left(\nu_{01}^2\, h_S^\dagger h_1 + \mathrm{h.c.}\right)+ \frac{1}{2} \left( \nu_{02}^2\, h_S^\dagger h_2 + \mathrm{h.c.} \right).
\end{aligned}
\end{equation}

In the case of real couplings, not all of the soft terms survive~\cite{Kuncinas:2020wrn}. This indicates that due to the minimisation conditions some of the bilinears could vanish. However, since we allow for complex quartic couplings it is natural to assume that the soft terms could also be promoted to complex parameters. Accounting for complex soft terms, depending on which terms are considered, there could be some freedom to rotate away phases of the soft terms as in the case of the quartic couplings, see eq.~\eqref{Eq:V4C_explicit_phases}. With additional soft terms CP violation would survive, as suggested by eq.~\eqref{Eq:Mu_Eq_Req}. Whenever there is a U(3) symmetry of the quadratic terms, CP is conserved.

{{\renewcommand{\arraystretch}{1.15}
\begin{table}[htb]
\caption{ Classification of the Yukawa Lagrangian charged under the $S_3$ symmetry, listing only cases which were not covered in Table~\ref{Table:CPV_models_list}. The presented cases require soft symmetry breaking of the scalar potential. In the last column models with complex vevs and which require complex Yukawa couplings are identified by $y_i^{(u,d)} \in \mathbb{C}$. Otherwise, it is sufficient to have real Yukawa couplings to generate the CKM matrix. Cases with real vevs require strictly complex Yukawa couplings.}
\label{Table:SSB}
\begin{center}
\begin{tabular}{|c|c|c|}\hline\hline
Vacuum & vevs & $\mathcal{L}_Y$ \\ \hline\hline
R-0 & $\left( 0,\,0,\,0 \right)$ & - \\ \hline
R-II-1a &  $\left( 0,\,w_2,\,w_S \right)$ & trivial\\ \hline
R-II-1b,c &  $\left( w_1,\, \pm w_1/\sqrt{3},\, w_S \right)$ & trivial  \\ \hline
R-II-2 &  $\left( 0,\, w_2,\,0 \right)$ & - \\ \hline
R-II-3 & $\left( w_1,\, w_2,\, 0 \right)$ & any except for trivial \\ \hline
R-III &  $\left( w_1,\, w_2,\, w_S \right)$ & any \\ \hline
R-III-s & $\left( w_1,\,0,\, w_S \right)$ & any \\ \hline\hline
C-III-b & $\left( \pm i w_1,\, 0 ,\, \hat w_S \right)$ & any with $y_i^{(u,d)} \in \mathbb{C}$  \\ \hline
C-III-c & $\left( \hat w_1 e^{i \sigma_1} ,\, \hat w_2 e^{i \sigma_2} ,\, 0\right)$ & any except for trivial \\ \hline
C-III-d & $\left( \pm i \hat w_1,\, \hat w_2 ,\, \hat w_S\right)$ & any with $y_i^{(u,d)} \in \mathbb{C}$  \\ \hline
C-III-e & $\left( \pm i \hat w_1,\, -\hat w_2 ,\, \hat w_S\right)$ & any with $y_i^{(u,d)} \in \mathbb{C}$  \\ \hline
C-III-f & $\left( \pm i \hat w_1,\, i \hat w_2 ,\, \hat w_S\right)$ &  any \\ \hline
C-III-g & $\left( \pm i \hat w_1,\, -i \hat w_2  ,\, \hat w_S\right)$ & any \\ \hline
C-III-i & $\left( \sqrt{\frac{3 \left( 1 + \tan^2 \sigma_1 \right)}{1 + 9 \tan^2 \sigma_1}} \hat w_2 e^{i \sigma_1} ,\, \pm \hat w_2 e^{-i \arctan(3 \tan \sigma_1)} ,\, \hat w_S\right)$ & any \\ \hline
C-IV-a & $\left( \hat w_1 e^{i \sigma_1} ,\, 0 ,\, \hat w_S\right)$ &  any \\ \hline
C-IV-b & $\left( \hat w_1,\, \pm i \hat w_2 ,\, \hat w_S\right)$ &  any \\ \hline
C-IV-d & $\left( \hat w_1 e^{i \sigma_1},\, \pm \hat w_2 e^{i \sigma_1} ,\, \hat w_S\right)$ & any \\ \hline
C-IV-e & $\left( \sqrt{-\frac{\sin 2 \sigma_2}{\sin 2 \sigma_1}} \hat w_2 e^{i \sigma_1},\, \hat w_2 e^{i \sigma_2} ,\, \hat w_S\right)$ & any \\ \hline
\end{tabular}
\end{center}
\end{table}}

Full analysis of the softly broken $S_3$-symmetric model is beyond the scope of the paper due to the many possibilities: there could be a single soft term, or two, or three or all four present. In Table~\ref{Table:CPV_models_list} ten cases were presented, which do not require soft terms. With the introduction of soft terms a few hundred different cases are possible. Now we have new cases, we list in Table~\ref{Table:SSB} the Yukawa Lagrangian structures one might consider. An interesting case is C-IV-b with vacuum given by $\left( \hat w_1,\, \pm i \hat w_2 ,\, \hat w_S\right)$. By performing an analysis of the Yukawa sector alone, we found that it is possible to generate a complex CKM matrix by means of the imaginary unit, with no need for an arbitrary phase.

\section{Discussion}\label{Sec:Discussion}

In total there are four $S_3$-based models which do not require complex quartic couplings but due to phases of vevs yield spontaneous CP violation. These are~\cite{Emmanuel-Costa:2016vej}: C-III-a, C-III-h, C-IV-c, C-IV-f. Two of these (C-III-a, C-III-h) require all of the fermions to be trivially charged under $S_3$, or in other words fermions couple only to $h_S$, the $S_3$ singlet and SU(2) scalar doublet. The C-III-a model was found to contain a viable dark matter candidate~\cite{Kuncinas:2022whn}. It would be interesting to see how the parameter space of the model changes when a complex coupling is introduced since C-III-a could be completely ruled out assuming some specific DM halo distribution profiles~\cite{Kuncinas:2023hpf}.

The other two models (C-IV-c, C-IV-f) do not survive when complex quartic couplings are introduced since $\lambda_4$ is not allowed to be complex in these cases. However this fact (also for the C-V case) does not contradict the results of Ref.~\cite{Branco:2015bfb} where it is conjectured that whenever a symmetry of the scalar potential prevents explicit CP violation it also prevents spontaneous CP violation, since here we are considering specific vacuum directions. Apart from that, both C-IV-c and C-IV-f could yield a good fit to the CKM matrix. However, both models suffer from an unrealistic scalar particle content---massless states. Soft symmetry breaking of the scalar potential could remove the massless scalars.

With complex quartic couplings there is more freedom to implement CP violation. First of all, models with real vevs (R-I-1, R-I-2a, R-I-2b,c) can result in explicit CP violation. However, only R-I-1 could yield a realistic Yukawa sector. Explicit CP violation is possible in C-I-a, C-III-a, C-III-h, C-IV-g, C-V. Out of these only C-I-a cannot account for the observed fermionic content. Neither C-III-a nor C-III-h is appealing since they require a trivial Yukawa sector and would not explain the source of CP violation required in the Yukawa Lagrangian. On the other hand, C-IV-g is capable of generating the experimentally observed CKM matrix with only a single phase, $\sigma_1$. This model is unique to the scalar potential with complex couplings. Unfortunately, this model is ruled out, unless soft symmetry breaking is introduced, due to unrealistic scalar masses.

As a consequence, the only ``good" model is C-V. It should be noted that although it might look like the most general one, it is not so since not all other models are contained within C-V. The complex CKM matrix is in this case generated by two phases, $\sigma_1$ and $\sigma_2$, coming from vevs. There is a viable region of parameter space with either one of the phases close to zero allowing to generate the correct complex CKM matrix. A remarkable aspect of the model is the possibility to have light neutral scalars at the $\mathcal{O}(\text{MeV})$ scale. It is possible to fit the scalar-fermion couplings in the C-V case in such a way that the $\mathcal{O}(\text{MeV})$ scalars escape detection. This is due to the large freedom that exists in the choice of the Yukawa Lagrangian. It remains to be seen whether such light scalars are not already ruled out by experiment due to some LHC channel which we did not take into account here. The second panel of Fig.~\ref{Fig:C-V_masses} illustrates the fact that in the C-V case the range of the lightest scalar mass can vary from a few MeV up to around 300 GeV.

The $S_3$-symmetric 3HDM has a very rich phenomenology, allowing for CP violation either explicit or spontaneous in a large variety of cases, depending on the region of parameter space. Different regions of parameter space correspond to different vacua with implications that were outlined in this paper.

\section*{Acknowledgements}
PO~is supported in part by the Research Council of Norway.
The work of AK and MNR was partially supported by Funda\c c\~ ao 
para  a  Ci\^ encia e a Tecnologia  (FCT, Portugal)  through  the  projects  
CFTP-FCT Unit UIDB/00777/2020 and UIDP/00777/2020, CERN/FIS-PAR/0002/2021 and CERN/FIS-PAR/0008/2019, which are  partially  funded  through
POCTI  (FEDER),  COMPETE,  QREN  and  EU. Furthermore, the work of AK has been supported by the FCT PhD fellowship with reference UI/BD/150735/2020.
We also thank the University of Bergen  and
CFTP/IST/University of Lisbon, where collaboration visits took place. 

\appendix 

\section{Reducible representation}\label{App:Derman_pot}

In the reducible-triplet framework the $S_3$ symmetry is manifest. The potential was written by Derman~\cite{Derman:1978rx},
\begin{subequations}
\begin{align}
V_2 &= -\lambda  \sum_i \left(\phi_i^\dagger \phi_i\right) + \frac{1}{2} \gamma \sum_{i<j}  \left[ \phi_i^\dagger \phi_j + \hc\right], \\
\begin{split}V_4 &=  A \sum_i\left( \phi_i^\dagger \phi_i \right)^2+\sum_{i<j} \left\lbrace C \left( \phi_i^\dagger \phi_i \right) \left( \phi_j^\dagger \phi_j \right) + \bar{C}\,\big| \phi_i^\dagger \phi_j \big|^2 + \frac{1}{2}D\left[ \left( \phi_i^\dagger \phi_j \right)^2 + \hc \right]
\right\rbrace\\
&\quad+\frac{1}{2}\sum_{i\neq j} \left[   E_1 \left( \phi_i^\dagger \phi_i \right)\left( \phi_i^\dagger \phi_j \right) + \hc\right] \\
&\quad+\sum_{i \neq j \neq k \neq i ,j<k} \left\lbrace \frac{1}{2} E_2 \left[ \left( \phi_i^\dagger \phi_j \right)\left( \phi_k^\dagger \phi_i \right)  + \hc\right] + \frac{1}{2} E_3 \left[ \left( \phi_i^\dagger \phi_i \right)\left( \phi_k^\dagger \phi_j + \hc\right)\right]\right.\\
&\hspace{86pt}\left. + \frac{1}{2}\left[ E_4  \left( \phi_i^\dagger \phi_j \right)\left( \phi_i^\dagger \phi_k \right) + \hc \right]
\right\rbrace.\end{split}
\end{align}
\end{subequations}
In total there are two complex couplings, $E_1$ and $E_4$, which, for simplicity, shall be written as
\begin{equation}
E_i \to e^{i \theta_i} E_i,\quad i=\{1,4\}.
\end{equation}

The reducible triplet fields are related to the irreducible representation~\eqref{Eq.Doublets_components} by transformations:
\begin{subequations}
\begin{alignat}{3}
& \textbf{2:}\quad{}&\begin{pmatrix}
h_1 \\ h_2 
\end{pmatrix} &= \begin{pmatrix}
\frac{1}{\sqrt{2}} \left( \phi_1 - \phi_2 \right)\\
\frac{1}{\sqrt{6}} \left( \phi_1 + \phi_2 - 2\phi_3 \right)
\end{pmatrix},\\
&\textbf{1:} {}&h_S &= \frac{1}{\sqrt{3}} \left( \phi_1 + \phi_2 + \phi_3 \right).
\end{alignat}
\end{subequations}
Using the above transformations we can relate couplings in different representations:
\begin{subequations}
\begin{align}
\mu_0^2 &= \gamma - \lambda,\\
\mu_1^2 &= - \left( \frac{1}{2} \gamma + \lambda\right),\\
\lambda_1 &= \frac{1}{12}\left( 4A + 4C + \bar{C} + D - 4 E_1 \cos \theta_1 + E_2 - 2E_3 + E_4 \cos \theta_4 \right),\\
\lambda_2 &=\frac{1}{4}\left( -\bar{C} + D + E_2 - E_4 \cos \theta_4 \right),\\
\lambda_3 &= \frac{1}{12} \left( 2 A - C + 2 \bar{C} + 2 D - 2 E_1 \cos \theta_1 - E_2 + 2 E_3 - E_4 \cos \theta_4 \right),\\
\begin{split}\lambda_4 &= \frac{1}{6 \sqrt{2}} \Big( 4 A - 2 C - 2 \bar{C} - 2 D - E_1 \cos \theta_1  + E_2 + E_3 +  E_4 \cos \theta_4\\&\hspace{50pt}- 3 i \left[  E_1 \sin \theta_1 - E_4 \sin \theta_4 \right] \Big),\end{split}\\
\lambda_5 &= \frac{1}{6}\left( 4 A +4 C -2\bar{C} - 2 D + 2 E_1 \cos \theta_1 - 2 E_2 + E_3 - 2 E_4 \cos \theta_4 \right),\\
\lambda_6 &= \frac{1}{6}\left( 4 A - 2 C + 4 \bar{C} - 2 D + 2 E_1 \cos \theta_1 + E_2 - 2 E_3 - 2 E_4 \cos \theta_4 \right),\\
\begin{split} \lambda_7 &= \frac{1}{12} \Big( 4 A - 2 C - 2 \bar{C} + 4 D + 2 E_1 \cos \theta_1  - 2 E_2 - 2 E_3 +  E_4 \cos \theta_4\\&\hspace{40pt}- 3 i \left[ 2 E_1 \sin \theta_1 + E_4 \sin \theta_4 \right] \Big),\end{split}\\
\lambda_8 &= \frac{1}{3} \left( A + C + \bar{C} + D + 2  E_1 \cos \theta_1 + E_2 + E_3 +  E_4 \cos \theta_4\right),
\end{align}
\end{subequations}
where $\{\lambda_4, \, \lambda_7\} \in \mathbb{C}$. In turn, the complex $\lambda_i$ can be split into a real and an imaginary part as $\lambda_i = \lambda_i^\mathrm{R} + i \lambda_i ^ \mathrm{I}$, which is the notation used throughout most of the article. The parts would be expressed as $\lambda_i^\mathrm{R} = f_1\left(A,\, C,\, \bar C,\, D,\, E_1,\, E_2,\, E_3,\, E_4\right)$ and $i\lambda_i^\mathrm{I} = if_2\left(E_1,\, E_4\right)$.

\section{Minimisation conditions}

Here, we present derivatives of the potential for different cases. The scalar potential was minimised with respect to the neutral fields $\eta_i$ and $\chi_i$.

\subsection{Minimisation conditions with phases \boldmath$\theta_i$ and \boldmath$\alpha_i$}\label{App:Derivatives_w_D_rot}

We start by considering the most general phase rotation of the $S_3$ doublet~\eqref{Eq.h_ph_theta} and phases of the couplings $\lambda_4$ and $\lambda_7$~\eqref{Eq.l4l7_alpha}. In this case we can simplify vevs and allow strictly for real configurations, $\{w_1,\,w_2,\,w_S\}$, due to the phase rotation of the doublets. The minimisation conditions can be written as
\begin{subequations}
\begin{align}
\begin{split} \dfrac{\partial V}{\partial \eta_1} \Bigg|_v =\,& \mu_1^2   w_1 
+\lambda_1   w_1 \left(   w_1^2 +   w_2^2 \right)
-2 \lambda_2 \sin^2(\theta_1-\theta_2)   w_1   w_2^2\\&
+\lambda_3 \left[   w_1^3 + \cos(2\theta_1-2\theta_2)   w_1   w_2^2 \right]\\&
+\lambda_4 \left[ \cos(2 \theta_1 - \theta_2 + \alpha_4)  + 2 \cos(\theta_2 + \alpha_4)\right]   w_1   w_2   w_S\\&
+\frac{1}{2} \left( \lambda_5 + \lambda_6 \right)  w_1   w_S^2 + \lambda_7 \cos(2\theta_1 + \alpha_7)  w_1  w_S^2,
\end{split}\\
\begin{split} \dfrac{\partial V}{\partial \eta_2} \Bigg|_v =\,& \mu_1^2 w_2 
+\lambda_1   w_2 \left(   w_1^2 +   w_2^2 \right)
-2 \lambda_2 \sin^2(\theta_1-\theta_2)   w_1^2   w_2\\&
+\lambda_3 \left[   w_2^3 + \cos(2\theta_1-2\theta_2)   w_1^2   w_2 \right]\\&
+\frac{1}{2} \lambda_4 \left\{ \left[ \cos(2 \theta_1 - \theta_2 + \alpha_4)  + 2 \cos(\theta_2 + \alpha_4)\right]w_1^2 - 3 \cos(\theta_2+\alpha_4) w_2^2 \right\}  w_S\\&
+\frac{1}{2} \left( \lambda_5 + \lambda_6 \right)  w_2   w_S^2 + \lambda_7 \cos(2\theta_2 + \alpha_7)  w_2  w_S^2,
\end{split}\\
\begin{split} \dfrac{\partial V}{\partial \eta_S} \Bigg|_v =\,&   \mu_0^2   w_S
+ \frac{1}{2} \lambda_4 \left\{ \left[ \cos(2 \theta_1 -\theta_2+\alpha_4) + 2 \cos(\theta_2+\alpha_4) \right] w_1^2  w_2 - \cos(\theta_2+\alpha_4)  w_2^3 \right\}\\&
+ \frac{1}{2} \left( \lambda_5 +\lambda_6 \right) \left(   w_1^2 +   w_2^2 \right)   w_S
+ \lambda_7 \left[ \cos(2 \theta_1+\alpha_7) w_1^2 + \cos(2 \theta_2+\alpha_7) w_2^2  \right]  w_S\\&
+ \lambda_8   w_S^3    ,\end{split}\\
\begin{split} \dfrac{\partial V}{\partial \chi_1} \Bigg|_v =\,&  
- \lambda_2 \sin(2\theta_1-2\theta_2) w_1  w_2^2
-\lambda_3 \sin(2\theta_1-2 \theta_2) w_1  w_2^2\\&
-\lambda_4 \sin(2\theta_1 -\theta_2+\alpha_4)   w_1   w_2   w_S
- \lambda_7 \sin(2\theta_1+\alpha_7)   w_1   w_S^2     ,\end{split}\\
\begin{split} \dfrac{\partial V}{\partial \chi_2} \Bigg|_v =\,&  
\lambda_2 \sin(2\theta_1-2\theta_2) w_1^2  w_2
+\lambda_3 \sin(2\theta_1-2 \theta_2) w_1^2  w_2\\&
+\frac{1}{2}\lambda_4 \left[ \sin(2\theta_1 -\theta_2+\alpha_4) w_1^2 - \sin(\theta_2+\alpha_4)\left( 2 w_1^2 - w_2^2 \right)\right]w_S\\&
- \lambda_7 \sin(2\theta_2+\alpha_7)   w_2   w_S^2     ,\end{split}\\
\begin{split} \dfrac{\partial V}{\partial \chi_S} \Bigg|_v =\,& 
\frac{1}{2} \lambda_4 \left\{ \left[ \sin(2 \theta_1 -\theta_2+\alpha_4) + 2 \sin(\theta_2+\alpha_4) \right] w_1^2 w_2 - \sin(\theta_2+\alpha_4) w_2^3 \right\}\\&
+ \lambda_7 \left[ \sin(2 \theta_1+\alpha_7) w_1^2 + \sin(2 \theta_2+\alpha_7) w_2^2  \right]  w_S.\end{split}
\end{align}
\end{subequations}

\subsection{Minimisation conditions for real vacua}\label{App:Derivatives_Rvev}

The most general real vacuum form yields the following first order derivatives:
\begin{subequations}
\begin{align}
 \dfrac{\partial V}{\partial \eta_1} \Bigg|_v ={}& \frac{1}{2} w_1 \left[ 2 \mu_1^2 + 2 \left(\lambda_1 + \lambda_3\right)\left(w_1^2+w_2^2\right) + 6 \lambda_4^\mathrm{R} w_2 w_S + \left(\lambda_5 + \lambda_6 + 2 \lambda_7\right) w_S^2\right],\\
\begin{split} \dfrac{\partial V}{\partial \eta_2} \Bigg|_v ={}& \frac{1}{2} \left[ 2 \mu_1^2 w_2 + 2 \left( \lambda_1 + \lambda_3 \right)\left( w_1^2 + w_2^2\right)w_2   +3 \lambda_4^\mathrm{R}\left(w_1^2 - w_2^2\right)w_S \right.\\&+ \left.\left( \lambda_5 + \lambda_6 + 2 \lambda_7 \right)w_2 w_S^2 \right],\end{split}\\
\begin{split} \dfrac{\partial V}{\partial \eta_S} \Bigg|_v ={}& \frac{1}{2} \left[ 2 \mu_0^2 w_S + \lambda_4^\mathrm{R} \left( 3 w_1^2 w_2 - w_2^3 \right)   + \left( \lambda_5 + \lambda_6 + 2 \lambda_7 \right)\left(w_1^2 + w_2^2\right)w_S \right. \\&+ \left. 2 \lambda_8 w_S^3\right],\end{split}\\
 \dfrac{\partial V}{\partial \chi_1} \Bigg|_v ={}& - \lambda_4^\mathrm{I} w_1 w_2 w_S,\label{Eq:DV_Dchi1}\\
 \dfrac{\partial V}{\partial \chi_2} \Bigg|_v ={}& -\frac{1}{2} \lambda_4^\mathrm{I} \left(w_1^2-w_2^2\right) w_S,\label{Eq:DV_Dchi2}\\
 \dfrac{\partial V}{\partial \chi_S} \Bigg|_v ={}& \frac{1}{2}\lambda_4^\mathrm{I}\left(3 w_1^2 - w_2^2\right)w_2.\label{Eq:DV_Dchi3}
\end{align}
\end{subequations}

\subsection{Minimisation conditions for complex vacua}\label{App:Derivatives_Cvev}

The most general complex vacuum form with a single complex coupling $\lambda_4$ yields the following first order derivatives:
\begin{subequations} \label{Eq:Derivatives_Cvev}
\begin{align}
\begin{split} \dfrac{\partial V}{\partial \eta_1} \Bigg|_v ={}& \mu_1^2 \cos\sigma_1 \hat w_1 
+\lambda_1 \cos\sigma_1 \hat w_1 \left( \hat w_1^2 + \hat w_2^2 \right)
+2 \lambda_2 \sin(\sigma_1-\sigma_2)\sin\sigma_2 \hat w_1 \hat w_2^2\\&
+\lambda_3 \left[ \cos\sigma_1 \hat w_1^3 + \cos(\sigma_1-2\sigma_2) \hat w_1 \hat w_2^2 \right]\\&
+\lambda_4^\mathrm{R} \left[ 3 \cos\sigma_1 \cos\sigma_2 + \sin\sigma_1 \sin\sigma_2 \right] \hat w_1 \hat w_2 \hat w_S\\&
-\lambda_4^\mathrm{I} \sin(\sigma_1 +\sigma_2) \hat w_1 \hat w_2 \hat w_S
+\frac{1}{2} \left( \lambda_5 + \lambda_6 + 2 \lambda_7 \right) \cos\sigma_1 \hat w_1 \hat w_S^2,
\end{split}\\
\begin{split} \dfrac{\partial V}{\partial \eta_2} \Bigg|_v ={}&  \mu_1^2 \cos\sigma_2 \hat w_2
+ \lambda_1 \cos\sigma_2 \hat w_2 \left( \hat w_1^2 + \hat w_2^2 \right)
- 2 \lambda_2 \sin\sigma_1 \sin(\sigma_1-\sigma_2) \hat w_1^2 \hat w_2\\&
+ \lambda_3 \left[ \cos(2 \sigma_1 -\sigma_2) \hat w_1^2 \hat w_2 + \cos\sigma_2 \hat w_2^3 \right]\\&
+ \frac{1}{2} \lambda_4^\mathrm{R} \left[ \left( 2 + \cos2 \sigma_1 \right)\hat w_1^2 - \left( 2 + \cos2 \sigma_2 \right) \hat w_2^2 \right] \hat w_S\\&
- \lambda_4^\mathrm{I} \left( \cos\sigma_1 \sin\sigma_1 \hat w_1^2 - \cos\sigma_2\sin\sigma_2 \hat w_2^2 \right) \hat w_S\\&
+ \frac{1}{2} \left( \lambda_5 +\lambda_6 +2 \lambda_7 \right) \cos\sigma_2 \hat w_2 \hat w_S^2,\end{split}\\
\begin{split} \dfrac{\partial V}{\partial \eta_S} \Bigg|_v ={}&   \mu_0^2 \hat w_S
+ \frac{1}{2} \lambda_4^\mathrm{R} \left\lbrace \left[ \cos(2 \sigma_1 -\sigma_2) + 2 \cos\sigma_2 \right] \hat w_1^2 \hat w_2 - \cos\sigma_2 \hat w_2^3 \right\rbrace\\&
- \frac{1}{2} \lambda_4^\mathrm{I} \left\lbrace \left[ \sin(2 \sigma_1 -\sigma_2) + 2 \sin\sigma_2 \right] \hat w_1^2 \hat w_2 - \sin\sigma_2 \hat w_2^3 \right\rbrace\\&
+ \frac{1}{2} \left( \lambda_5 +\lambda_6 \right) \left( \hat w_1^2 + \hat w_2^2 \right) \hat w_S
+ \lambda_7 \left( \cos2 \sigma_1 \hat w_1^2 + \cos2 \sigma_2 \hat w_2^2  \right) \hat w_S + \lambda_8 \hat w_S^3    ,\end{split}\\
\begin{split} \dfrac{\partial V}{\partial \chi_1} \Bigg|_v ={}&  \mu_1^2 \sin\sigma_1 \hat w_1 
+\lambda_1 \sin\sigma_1 \hat w_1 \left( \hat w_1^2 + \hat w_2^2 \right)
- 2 \lambda_2 \sin(\sigma_1-\sigma_2)\cos\sigma_2 \hat w_1 \hat w_2^2\\&
+\lambda_3 \left[ \sin\sigma_1 \hat w_1^3 - \sin(\sigma_1-2 \sigma_2) \hat w_1 \hat w_2^2 \right]
+\lambda_4^\mathrm{R} \sin(\sigma_1 +\sigma_2) \hat w_1 \hat w_2 \hat w_S\\&
-\lambda_4^\mathrm{I} \left[ 3 \sin\sigma_1 \sin\sigma_2 + \cos\sigma_1 \cos\sigma_2 \right] \hat w_1 \hat w_2 \hat w_S\\&
+\frac{1}{2} \left( \lambda_5 + \lambda_6 - 2 \lambda_7 \right) \sin\sigma_1 \hat w_1 \hat w_S^2,\end{split}\\
\begin{split} \dfrac{\partial V}{\partial \chi_2} \Bigg|_v ={}& \mu_1^2 \sin\sigma_2 \hat w_2 
+\lambda_1 \sin\sigma_2 \hat w_2 \left( \hat w_1^2 + \hat w_2^2 \right)
+2 \lambda_2 \sin(\sigma_1-\sigma_2)\cos\sigma_1 \hat w_1^2 \hat w_2\\&
+\lambda_3 \left[ \sin(2 \sigma_1- \sigma_2) \hat w_1^2 \hat w_2 + \sin\sigma_2 \hat w_2^3 \right]\\&
+\lambda_4^\mathrm{R} \left[ \cos\sigma_1 \sin\sigma_1 \hat w_1^2 - \cos\sigma_2 \sin\sigma_2 \hat w_2^2 \right]\hat w_S \\&
- \frac{1}{2} \lambda_4^\mathrm{I} \left[ \left( 2 - \cos2 \sigma_1 \right) \hat w_1^2 - \left( 2 - \cos2 \sigma_2 \right)\hat w_2^2   \right] \hat w_S\\&
+\frac{1}{2} \left( \lambda_5 + \lambda_6 - 2 \lambda_7 \right) \sin\sigma_2 \hat w_2 \hat w_S^2,\end{split}\\
\begin{split} \dfrac{\partial V}{\partial \chi_S} \Bigg|_v ={}&
\frac{1}{2} \lambda_4^\mathrm{R} \left\lbrace \left[ \sin(2 \sigma_1 -\sigma_2) + 2 \sin\sigma_2 \right] \hat w_1^2 \hat w_2 - \sin\sigma_2 \hat w_2^3 \right\rbrace\\&
+ \frac{1}{2} \lambda_4^\mathrm{I} \left\lbrace \left[ \cos(2 \sigma_1 -\sigma_2) + 2 \cos\sigma_2 \right] \hat w_1^2 \hat w_2 - \cos\sigma_2 \hat w_2^3 \right\rbrace\\&
+ \lambda_7 \left( \sin2 \sigma_1 \hat w_1^2 + \sin2 \sigma_2 \hat w_2^2  \right) \hat w_S.\end{split}
\end{align}
\end{subequations}

\section{CP-odd invariants}\label{App:CPodd_inv}

For each order of the $Z$-tensors we shall present only a sufficient number of invariants and not all possible unique ones. There are no CP-odd invariants containing less than five $Z$-tensors.

\subsection{CP-odd invariants containing five \boldmath$Z$-tensors}

There are only two independent CP-odd invariants containing five $Z$-tensors. They are given by
\begin{subequations}
\begin{align}
\begin{split}
\mathrm{I}_{5Z}^{(1)}={}& \mathbb{I}\mathrm{m} \left[Z_{aabc}Z_{dbef}Z_{cghe}Z_{idgh}Z_{fijj}\right]\\
={}& 16 \lambda_4^\mathrm{R} \lambda_4^\mathrm{I} \lambda_7 (4\lambda_1-\lambda_5-2\lambda_8)^2,\end{split}\\
\begin{split}
\mathrm{I}_{5Z}^{(2)}={}& \mathbb{I}\mathrm{m} \left[Z_{abbc}Z_{daef}Z_{cghe}Z_{idgh}Z_{fjji}\right] \\
={}& 16 \lambda_4^\mathrm{R} \lambda_4^\mathrm{I} \lambda_7(2\lambda_1-2\lambda_2+4\lambda_3-\lambda_6-2\lambda_8)^2.\end{split}
\end{align}
\end{subequations}

There are four distinct ways these two invariants can vanish:
\begin{itemize}
\item {\bf Solution 0:} $\lambda_4^\mathrm{I}=0$.\\
In this case there is no explicit CP violation. The only possibility for CP violation depends on the vacuum. We may at most have spontaneous CP violation.  This case was covered previously \cite{Emmanuel-Costa:2016vej}.

\item {\bf Solution 1:} $\lambda_4^\mathrm{R}=0$.\\
In this case there is no explicit CP violation. We may perform a change of basis by $h_2\to i h_2$. This makes the whole potential real. The only possibility for CP violation depends on the vacuum. We may at most have spontaneous CP violation.  

\item {\bf Solution 2:} $\lambda_7=0$.\\
In this case we see from eq.~\eqref{Eq:V4C_explicit_phases} that the choice of $\theta=-\alpha_4$ results in a real potential. Or else, we could perform a rotation $\left( h_S \right)_\mathrm{old} \to e^{i \arg(\lambda_4)} \left( h_S^\prime \right)_\mathrm{new} $, where the argument of the complex $\lambda_4$ would cancel the phase $\alpha_4$, so that $\left(\lambda_4\right)_\mathrm{old} \to |\lambda_4|_\mathrm{new}$. In this case, the only possibility for CP violation depends on the vacuum. We may at most have spontaneous CP violation.  

\item {\bf Solution 3:} $\lambda_5=2(2\lambda_1-\lambda_8)$, $\lambda_6=2(\lambda_1-\lambda_2+2\lambda_3-\lambda_8)$ and $\lambda_4^\mathrm{R} \lambda_4^\mathrm{I} \lambda_7 \neq 0$.\\
These constraints are not sufficient to make $V_4$ CP-invariant. This is because there are still non-zero invariants containing six or more $Z$-tensors. We shall be imposing this condition when discussing higher-order CP-odd invariants.
\end{itemize}

\subsection{CP-odd invariants containing six \boldmath$Z$-tensors}

Two independent CP-odd invariants containing six $Z$-tensors are found:
\begin{subequations}
\begin{align}
\begin{split}
\mathrm{I}_{6Z}^{(1)}={}& \mathbb{I}\mathrm{m}\left[Z_{abcd}Z_{baef}Z_{gchi}Z_{djke}Z_{fkil}Z_{jglh}\right]\\
={}& -192 \lambda_4^\mathrm{R} \lambda_4^\mathrm{I} \lambda_7 (\lambda_1-\lambda_2-\lambda_8)\\
&\hspace{70pt}\times \left[|\lambda_4|^2 - 2 \lambda_7^2  - 2 \left( \lambda_1 - \lambda_2 - \lambda_8 \right)
\left( 3 \lambda_1 + \lambda_2 + 4 \lambda_3 - 3 \lambda_8 \right)  \right],
\end{split}\\
\begin{split}
\mathrm{I}_{6Z}^{(2)}={}& \mathbb{I}\mathrm{m}\left[Z_{abcd}Z_{baef}Z_{gchi}Z_{dejk}Z_{fhkl}Z_{lgij}\right]\\
={}&-192 \lambda_4^\mathrm{R} \lambda_4^\mathrm{I} \lambda_7 
\Big[|\lambda_4|^2 \left( 3 \lambda_1 + \lambda_2 + 4 \lambda_3 - 3 \lambda_8 \right) 
- 2 \lambda_7^2 \left( \lambda_4^\mathrm{R} - \lambda_4^\mathrm{I}  \right)\\
& \hspace*{17.5pt}- 4 \left( \lambda_1 + \lambda_3 - \lambda_8 \right)\\
& \hspace*{35pt} \times \left( 3 \lambda_1^2 + \lambda_2^2 + 6 \lambda_1 \lambda_3 + 2 \lambda_2 \lambda_3 + 4 \lambda_3^2 - \lambda_7^2 - 6 \lambda_1 \lambda_8 - 6 \lambda_3 \lambda_8 + 3 \lambda_8^2 \right) \Big].
\end{split}
\end{align}
\end{subequations}

These invariants both vanish in two cases:
\begin{itemize}
\item {\bf Solution 3.1:}
\begin{subequations}
\begin{align}
\lambda_5 ={}&2(\lambda_1+\lambda_2),\\
\lambda_6 ={}&4\lambda_3,\\
\lambda_8 ={}&\lambda_1-\lambda_2,\\
(\lambda_4^\mathrm{R})^2={}&\frac{\left[2\lambda_2+2\lambda_3+\lambda_7 \right]\left[2 (\lambda_2+\lambda_3)(2\lambda_2+2\lambda_3-\lambda_7)-(\lambda_4^\mathrm{I})^2\right]}{2\lambda_2+2\lambda_3-\lambda_7}.
\end{align}
\end{subequations}
Since $\lambda_4^\mathrm{R}=0$ would result in no explicit CP violation we can check for special cases:
\begin{subequations}
\begin{align}
(\lambda_4^\mathrm{I})^2 ={}& 2 \left( \lambda_2 + \lambda_3 \right) \left( 2 \lambda_2 + 2 \lambda_3 - \lambda_7 \right),\\
\lambda_7 ={}& - 2 \left(\lambda_2 + \lambda_3\right).
\end{align}
\end{subequations}

\item {\bf Solution 3.2:}
\begin{subequations}
\begin{align}
\lambda_5 ={}&2(2\lambda_1 - \lambda_8),\\
\lambda_6 ={}&2(\lambda_1 - \lambda_2 + 2\lambda_3 - \lambda_8	),\\
\begin{split}
(\lambda_4^\mathrm{R})^2 ={}&\frac{1}{2\lambda_7}(\lambda_1+\lambda_2+2 \lambda_3+\lambda_7-\lambda_8) \\
	&\times\Big[3 \lambda_1^2-3 \lambda_1 (2 \lambda_2-\lambda_7+2 \lambda_8)-\lambda_2^2-\lambda_2 (8 \lambda_3+\lambda_7-6 \lambda_8)\\
	&\hspace*{20pt}-4 \lambda_3^2+2 \lambda_3 \lambda_7+2 \lambda_7^2-3 \lambda_7 \lambda_8+3 \lambda_8^2\Big], \end{split}
	\label{Eq:lam4_R_sq}\\
\begin{split}
(\lambda_4^\mathrm{I})^2 ={}&-\frac{1}{2\lambda_7}(\lambda_1+\lambda_2+2 \lambda_3-\lambda_7-\lambda_8) \\
	&\times \Big[3 \lambda_1^2-3 \lambda_1 (2 \lambda_2+\lambda_7+2 \lambda_8)-\lambda_2^2-\lambda_2 (8 \lambda_3-\lambda_7-6 \lambda_8)\\
	&\hspace*{20pt}-4 \lambda_3^2-2 \lambda_3 \lambda_7+2 \lambda_7^2+3 \lambda_7 \lambda_8+3 \lambda_8^2\Big]. \end{split}
\end{align}
\end{subequations}
We note that the condition on $(\lambda_4^\mathrm{I})^2$ can be obtained from that on $(\lambda_4^\mathrm{R})^2$ by the substitution $\lambda_7\to-\lambda_7$. 
For future reference, we note that
\begin{equation}
|\lambda_4|^2=(\lambda_4^\mathrm{R})^2+(\lambda_4^\mathrm{I})^2
=2\lambda_7^2+2(\lambda_1-\lambda_2-\lambda_8)(3\lambda_1+\lambda_2+4\lambda_3-3\lambda_8).
\end{equation}
Moreover, since $\lambda_4^\mathrm{R} \lambda_4^\mathrm{I} \lambda_7 = 0$ results in no explicit CP violation we could check for the special case $\lambda_4^\mathrm{R} \lambda_4^\mathrm{I} = 0$, with the $\lambda_4$ coupling expressed in terms of the other $\lambda$s. Here, $\lambda_4^\mathrm{R}$ of eq.~(\ref{Eq:lam4_R_sq}) vanishes when
\begin{subequations}
\begin{align}
\begin{split} \lambda_2 + \lambda_3 ={}& \frac{1}{4} \Bigg[ \lambda_7 \pm \sqrt{3}\,\Big(4 \lambda_1^2 + 4\lambda_2^2 - 4 \lambda_2 \lambda_7 + 8 \lambda_2 \lambda_8 + 3 \lambda_7^2 \\
&\hspace*{65pt}- 4 \lambda_1\left( 2 \lambda_2 - \lambda_7 + 2\lambda_8 \right)  - 4 \lambda_7 \lambda_8 + 4 \lambda_8^2 \Big)^{1/2} \Bigg], \end{split}
\end{align}
and $\lambda_4^\mathrm{I}$ of eq.~(\ref{Eq:lam4_R_sq}) vanishes when in the above equation we replace $\lambda_7$ by $-\lambda_7$. Furthermore, either one vanishes when
\begin{equation}
\lambda_8 = \lambda_1 + \lambda_2 + 2 \lambda_3 \pm \lambda_7 \label{Eq:Sol3.2_l8}.
\end{equation}
\end{subequations}
\end{itemize}

\subsection{CP-odd invariants containing seven \boldmath$Z$-tensors}\label{Sect:sevenZs}

We find that it is sufficient to check a single CP-odd invariant,
\begin{equation}
\mathrm{I}_{7Z} = \mathbb{I}\mathrm{m}\left[Z_{abcd}Z_{eafc}Z_{bgdh}Z_{iejk}Z_{gflm}Z_{hlkn}Z_{minj}\right].
\end{equation}
Solution 3.1 yields:
\begin{equation}\label{Eq:Z7sol_1}
\mathrm{I}_{7Z} = 128 \lambda_4^\mathrm{R} \lambda_4^\mathrm{I} \lambda_7 \left[ \left( \lambda_4^\mathrm{R} \right)^2 + \left( \lambda_4^\mathrm{I} \right)^2  - 2 \lambda_7^2 \right]^2.
\end{equation}
After substituting for $\lambda_4^\mathrm{R}$ we find that the $\mathrm{I}_{7Z}$ CP invariant vanishes when
\begin{equation}
(\lambda_4^\mathrm{I})^2 =\frac{\left( \lambda_2 + \lambda_3 + \lambda_7 \right)\left( 2 \lambda_2 + 2 \lambda_3 - \lambda_7 \right)^2}{\lambda_7}.
\end{equation}
For Solution 3.2 we get the following expression:
\begin{equation}\label{Eq:Z7sol_2}
\begin{aligned}
\mathrm{I}_{7Z} ={}& 2304\, \lambda_4^\mathrm{R} \lambda_4^\mathrm{I} \lambda_7
(\lambda_1 - \lambda_2 - \lambda_8)^2 \\
&\times(\lambda_1 + \lambda_2 + 2\lambda_3 + \lambda_7 - \lambda_8)
(\lambda_1 + \lambda_2 + 2\lambda_3 - \lambda_7 - \lambda_8).
\end{aligned}
\end{equation}
Since eq.~\eqref{Eq:Sol3.2_l8} would result in CP conservation the only other possible condition for the CP invariant to vanish is
\begin{equation}
\lambda_8 = \lambda_1 - \lambda_2.
\end{equation}

The two cases mentioned above coincide as both require (we introduced $\lambda_{23} \equiv \lambda_2 + \lambda_3$):
\begin{subequations}\label{Eq:Sol_CP_Z7}
\begin{align}
\left( \lambda_4^\mathrm{R} \right)^2={}& -\frac{(\lambda_{23}-\lambda_7)(2\lambda_{23}+\lambda_7)^2}{\lambda_7},\\
\left( \lambda_4^\mathrm{I} \right)^2={}& \frac{(\lambda_{23}+\lambda_7)(2\lambda_{23}-\lambda_7)^2}{\lambda_7},\\
\lambda_4^\mathrm{R} \lambda_4^\mathrm{I} \lambda_7  \neq{}& 0,\\
\lambda_5 ={}& 2 \left( \lambda_1 + \lambda_2 \right),\\
\lambda_6 ={}& 4 \lambda_3,\\
\lambda_8 ={}& \lambda_1 - \lambda_2,
\end{align}
\end{subequations}
with
\begin{equation} \label{Eq:lam4-mod}
|\lambda_4|^2=(\lambda_4^\mathrm{R})^2+(\lambda_4^\mathrm{I})^2=2\lambda_7^2.
\end{equation}

If we prove that the scalar potential, specifically the quartic part, is CP-invariant provided that the eq.~\eqref{Eq:Sol_CP_Z7} constraints are satisfied, we would be assured that all other invariants would vanish. The phase-sensitive part of the quartic potential, (\ref{Eq:V4C_explicit_phases-0}), is given by
\begin{equation}
\begin{aligned}
V_4^\mathrm{phase} ={} &
\frac{\lambda_{23}}{2}\Big[(h_1^\dagger h_1 - h_2^\dagger h_2)^2+2(h_1^\dagger h_2)^2+2(h_2^\dagger h_1)^2+2(h_S^\dagger h_S)(h_1^\dagger h_1 + h_2^\dagger h_2)\\
&\hspace*{25pt}-(h_S^\dagger h_S)^2+4(h_1^\dagger h_S)(h_S^\dagger h_1)+4(h_2^\dagger h_S)(h_S^\dagger h_2)\Big]\\
&+ \left\lbrace \lambda_4\left[ (h_S^\dagger h_1)(h_1^\dagger h_2+h_2^\dagger h_1)+(h_S^\dagger h_2)(h_1^\dagger h_1-h_2^\dagger h_2) \right] + \hc \right\rbrace \\
&+ \left\lbrace\lambda_7 \left[(h_S^\dagger h_1)(h_S^\dagger h_1) + (h_S^\dagger h_2)(h_S^\dagger h_2)\right] +\hc\right\rbrace.
\end{aligned}
\end{equation}
Note that, apart from sign ambiguities, $\lambda_4$ is determined by $\lambda_{23}$ and $\lambda_7$.

We do not need to consider all combinations of $\lambda_{23}$ and $\lambda_7$. This is because $(\lambda_4^\mathrm{R})^2$ and $(\lambda_4^\mathrm{I})^2$  must necessarily be positive by definition. Moreover, since $\lambda_4$ is related to $\lambda_{23}$ and $\lambda_7$ by eq.~\eqref{Eq:Sol_CP_Z7}, parts of the $(\lambda_{23}-\lambda_7)$ plane are irrelevant for this analysis. The relevant regions consist of $\lambda_{23}=0$ (no restrictions on $\lambda_7$ apart from $\lambda_7 \neq 0$) as well as the region given by $\lambda_7^2-\lambda_{23}^2>0$, except for the two lines given by $\lambda_7=\pm\lambda_{23}$.

We found that when $\lambda_{23}=0$ a rotation into a new basis forces all couplings to become real. Such rotation is given by
\begin{equation}
\begin{pmatrix}
h_1 \\
h_2 \\
h_S
\end{pmatrix}
= \begin{pmatrix}
1 & 0 & 0 \\
0 & e^{i \theta} \cos \alpha & i e^{i \theta} \sin \alpha \\
0 & \sin \alpha & -i \cos \alpha 
\end{pmatrix} 
\begin{pmatrix}
h_1^\prime \\
h_2^\prime \\
h_S^\prime
\end{pmatrix},
\end{equation}
where
\begin{equation}
\theta ={} - \frac{\pi}{4},\qquad
\alpha ={} -\frac{1}{2} \arctan \sqrt{2}.
\end{equation}

For the remaining relevant part of the $(\lambda_{23}-\lambda_7)$ plane we utilise a basis transformation into a new basis with real coefficients, given by
\begin{equation}
\begin{pmatrix}
h_1 \\
h_2 \\
h_S
\end{pmatrix}  = 
\begin{pmatrix}
1 & 0 & 0 \\
0 & e^{i \theta} \cos \alpha &  e^{i \left( \theta + \phi \right)} \sin \alpha \\
0 & \sin \alpha & -e^{i \phi} \cos \alpha 
\end{pmatrix} 
\begin{pmatrix}
h_1^\prime \\
h_2^\prime \\
h_S^\prime
\end{pmatrix},
\end{equation}
where
\begin{subequations}
\begin{align}
\alpha &= \arctan\left(\frac{\lambda_{23}\sin2\theta}{\lambda_4^\mathrm{I}\cos\theta-\lambda_4^\mathrm{R}\sin\theta}\right),\\
\phi &= \arctan\left( \frac{\lambda_4^\mathrm{I} \cos \theta + \lambda_4^\mathrm{R} \sin \theta }{\lambda_{23}\sin\left(2 \alpha\right) + \left( \lambda_4^\mathrm{R} \cos \theta - \lambda_4^\mathrm{I} \sin \theta \right)\cos\left(2 \alpha\right) } \right),
\end{align}
\end{subequations}
while the general form of the $\theta$ angle is given by
\begin{equation}
2 \theta = \pm \arctan\left(\frac{\sqrt{3}\lambda_7+2\sqrt{\lambda_7^2-\lambda_{23}^2}}{\lambda_{23}}\right) + \left\lbrace \begin{array}{ll}
& \pi \text{ when } \lambda_{23}\lambda_7<0,\\
& 0 \text{ else}.
\end{array}\right.
\end{equation}

To sum up, there always exists a basis with real coefficients if the conditions of eq.~\eqref{Eq:Sol_CP_Z7} are satisfied. Therefore, there will in this case not be explicit CP violation in the scalar sector. Furthermore, we checked for CP-odd invariants containing eight $Z$-tensors. All of these tensors vanish provided that conditions~\eqref{Eq:Sol_CP_Z7} are satisfied. 

\subsection{CP-odd invariants containing both \boldmath$Y$- and \boldmath$Z$-tensors} 

For the trivial solutions (Solution 0, 1, 2) we get that the whole potential is CP invariant. However, Solution 3 might result in CP violation coming from the $V_2$ scalar potential part. Therefore, one might need to impose additional constraints involving the bilinear couplings $\mu_0^2$ and $\mu_1^2$. The lowest order CP-odd invariant containing a mixture of $Y$- and $Z$-tensors is given by
\begin{equation}\label{Eq:Mu_Eq_Req}
\mathrm{I}_{2Y3Z}=\mathbb{I}\mathrm{m} \left[Z_{abcd}Z_{befg}Z_{dchf}Y_{ga}Y_{eh} \right] = 16 \lambda_4^\mathrm{R} \lambda_4^\mathrm{I} \lambda_7(\mu_0^2-\mu_1^2)^2.
\end{equation}
The quadratic scalar potential part $V_2$ becomes insensitive to basis changes, invariant under U(3), if $\mu_1^2=\mu_0^2$ or one of these quartic couplings vanishes.

\section{Basis transformation for C-IV-e}\label{App:BT-complex}

For the particular vacuum $(\pm \hat w_1e^{-i\arctan(3\tan\sigma_2)}, \sqrt{\frac{3(1+\tan^2\sigma_2)}{1+9\tan^2\sigma_2}}\hat w_1e^{i\sigma_2},w_S)$ discussed in section~\ref{Sec:Diff_Im_Models}, the following basis transformation yields both a real potential and a real vacuum:
\begin{equation}
\begin{pmatrix}
	h_1\\
	h_2\\
	h_S
\end{pmatrix}
=
\begin{pmatrix}
	e^{-i\gamma_0}\cos\alpha & e^{-i(\gamma_0+\gamma_2)}\sin\alpha & 0\\
	e^{-i(\gamma_0-\gamma_1)}\sin\alpha & -e^{-i(\gamma_0-\gamma_1+\gamma_2)}\cos\alpha & 0\\
	0 & 0 & 1
\end{pmatrix}
\begin{pmatrix}
	h_1^\prime\\
	h_2^\prime\\
	h_S^\prime
\end{pmatrix},
\end{equation}
with
\begin{subequations}
\begin{align}
\gamma_0&=\arctan\left(\frac{
\sqrt{3}(1+2\cos2\alpha)}
{\sqrt{-1-2\cos4\alpha}}
\right),\\
\gamma_1&=-\arctan\left(\frac{\sqrt{-1-2\cos4\alpha}}
{\sqrt{3}\cos2\alpha}
\right),\\
\gamma_2&=\arctan\left(\frac{
	\sqrt{3}}
{\sqrt{-1-2\cos4\alpha}}
\right),
\end{align}
\end{subequations}
where $\alpha=5\pi/24$ if $\cos\sigma<0$ and $\alpha=7\pi/24$ if $\cos\sigma>0$ for $( \hat w_1e^{-i\arctan(3\tan\sigma)},\,\dots)$. For $(- \hat w_1e^{-i\arctan(3\tan\sigma)},\,\dots)$ the values are $\alpha=5\pi/24$ if $\cos\sigma>0$ and $\alpha=7\pi/24$ if $\cos\sigma<0$. It should be noted that solutions provided in terms of $\alpha$ are not unique, and there are several continuous regions which yield a rotation into a real basis. The chosen $\alpha$ values were picked as an illustrative example.

\section{Unitarity constraints}\label{App:Unitarity}

The study of unitarity in two-body scattering processes involving longitudinally polarized bosons and the Higgs boson in the SM were pioneered by Lee, Quigg and Thacker \cite{Lee:1977eg} and further on analysed in Refs.~\cite{Luscher:1988gc,Marciano:1989ns}. The process of finding the unitarity limit is straightforward due to the Goldstone equivalence theorem, which relates the longitudinally polarised vector boson and the Goldstone bosons in the high-energy limit \cite{Cornwall:1974km,Vayonakis:1976vz}. It is sufficient to consider the $2 \to 2 $ scattering processes of the scalar eigenstates. Some of these processes will be forbidden by the CP and $S_3$ symmetries. This indicates that there exists a basis with block-diagonal entries. This, in turn, would simplify computation of the eigenvalues. 

The unitarity conditions for the real $S_3$-symmetric 3HDM were derived earlier in Ref.~\cite{Das:2014fea}. Due to the $\lambda_4^\mathrm{I}$ term there will arise additional off-diagonal mixing terms in the $S$-matrix. The eigenvalues, which are common in both real, $\lambda_4^\mathrm{I}=0$, and complex models are
\begin{subequations}
\begin{align}
e_1 \,(b_6) & = \lambda_5 - \lambda_6,\\
e_2 \,(b_3) & = 2 \left( \lambda_1 - 5 \lambda_2 - 2 \lambda_3 \right),\\
e_3 \,(b_4,b_5) & = 2 \left( \lambda_1 \pm \lambda_2 - 2 \lambda_3 \right),\\
e_4 \,(a_2^\pm) & = \lambda_1+\lambda_2+2 \lambda_3+\lambda_8\pm \sqrt{(\lambda_1+\lambda_2+2 \lambda_3-\lambda_8)^2+8 \lambda_7^2},\\
e_5 \,(a_3^\pm)& = \lambda_1-\lambda_2+2 \lambda_3+\lambda_8\pm \sqrt{(-\lambda_1+\lambda_2-2 \lambda_3+\lambda_8)^2+2 \lambda_6^2},\\
e_6 \,(a_5^\pm)& = 5 \lambda_1-\lambda_2+2 \lambda_3+3 \lambda_8\pm \sqrt{(-5 \lambda_1+\lambda_2-2 \lambda_3+3 \lambda_8)^2+2 (2 \lambda_5+\lambda_6)^2},\\
e_7 \,(a_1^\pm)& = \frac{1}{2} \left(2 \lambda_1-2 \lambda_2+\lambda_5+\lambda_6\pm \sqrt{(-2 \lambda_1+2 \lambda_2+\lambda_5+\lambda_6)^2+16 |\lambda_4|^2 }\right).
\end{align}
\end{subequations}
In the notation of Ref.~\cite{Das:2014fea} these are $\{a_1^\pm,\, a_2^\pm,\, a_3^\pm,\, a_5^\pm,\, b_3,\, b_4,\, b_5,\, b_6\}$.

There are also additional eigenvalues. There is no simple way to write these as analytic expressions; they should be determined numerically from:
\begin{subequations}\label{Eq:S_eigen_to_solve}
\begin{align}
\mathcal{N}_{55} ={}& \begin{pmatrix}
2 \left( \lambda_1 + \lambda_2 \right) & -2i \lambda_4^\mathrm{I} & 2\lambda_4^\mathrm{R}\\
2i\lambda_4^\mathrm{I} & \lambda_5 - 2 \lambda_7 & 0 \\
2\lambda_4^\mathrm{R} & 0 & \lambda_5 + 2 \lambda_7
\end{pmatrix},\\
\mathcal{N}_{66} ={}& \begin{pmatrix}
2 \left( \lambda_1 + \lambda_2 + 4 \lambda_3 \right) & -6i \lambda_4^\mathrm{I} & 6\lambda_4^\mathrm{R}\\
6i\lambda_4^\mathrm{I} & \lambda_5 + 2 \lambda_6 - 6 \lambda_7 & 0 \\
6\lambda_4^\mathrm{R} & 0 & \lambda_5 + 2 \lambda_6 + 6 \lambda_7
\end{pmatrix}.
\end{align}
\end{subequations}
These eigenvalues simplify to $\mathcal{N}_{55}\,(b_2,\, a_4^\pm)$ and $\mathcal{N}_{66}\,(b_1,\, a_6^\pm)$ in the limit of $\lambda_4^\mathrm{I} = 0$.
A more detailed picture is discussed below.

\subsection{Deriving unitarity constraints}

The most trivial way to split the scattering matrix into a block-diagonal form is to group two-body states based on their charges,
\begin{equation}
S= \mathrm{diag}\left( S^0,\, S^+,\, S^{++} \right),
\end{equation} 
where the superscripts indicate if the $S$-matrix is neutral, singly charged or doubly charged. The scattering matrix $S_{i}=\left\langle  \Psi_i^n \,|\, \Psi_i^n\right\rangle$ is constructed from the two-particle states $\Psi_i^n$. These states can be read off directly from the quartic part of the scalar potential, which is expanded in terms of SU(2) doublets
\begin{equation}
h_i = \begin{pmatrix}
w_i^+ \\
n_i
\end{pmatrix}.
\end{equation}

When a two-particle state is constructed from identical species it gets a factor $1/\sqrt{2}$ due to the Bose-Einstein statistics. Finally, due to symmetries of the scalar potential, some of the off-diagonal matrix entries will be symmetric. In light of this it is useful to apply a unitary transformation
\begin{equation}
U = \frac{1}{\sqrt{2}} \begin{pmatrix}
1 & -1 \\
1 & 1
\end{pmatrix},
\end{equation}
which allows for removal of several entries in the $S$-matrix. Also, we shall be using identity matrices of dimension $n$ denoted as $\mathcal{I}_n$ and null matrices $\mathcal{O}_{m \times n}$ of dimension $m \times n$.

We start by considering the neutral two-body interactions. Exploiting the $\mathbb{Z}_2$ symmetry of $h_1$, we can split the states into three blocks:
\begin{subequations}
\begin{align}
\Psi^0_1 &= \Big\{ \left.| n_1 n_1\right\rangle,\, \left.| n_2 n_2\right\rangle,\, \left.| n_S n_S\right\rangle,\, \left.| n_2 n_S\right\rangle \Big\},\\
\Psi^0_2 &= \Big\{ \left.| n_1 n_2\right\rangle,\, \left.| n_1 n_S\right\rangle \Big\},\\
\Psi^0_3 &= \Big\{  \left.| w^+_1 w^-_1 \right\rangle,\ \left.| w^+_2 w^-_2 \right\rangle,\ \left.| w^+_2 w^-_S \right\rangle,\ \left.| w^+_S w^-_2 \right\rangle,\ \left.| n_1 n_1^\ast\right\rangle,  \\
&\qquad \left.| n_2 n_2^\ast \right\rangle,\ \left.| n_2 n_S^\ast \right\rangle,\ \left.| n_S n_2^\ast \right\rangle,\ \left.| w^+_S w^-_S \right\rangle,\ \left.| n_S n_S^\ast \right\rangle \Big\},\\
\Psi^0_4 &= \Big\{  \left.| w^+_1 w^-_2 \right\rangle,\ \left.| w^+_2 w^-_1 \right\rangle,\ \left.| w^+_1 w^-_S \right\rangle,\ \left.| w^+_S w^-_1 \right\rangle,\\
&\qquad  \left.| n_1 n_2^\ast \right\rangle,\ \left.| n_2 n_1^\ast \right\rangle,\ \left.| n_1 n_S^\ast \right\rangle,\ \left.| n_S n_1^\ast \right\rangle \Big\},
\end{align}
\end{subequations}
where the first two of the four $S$-matrices are
\begin{subequations}
\begin{align}
S_1^0 & = \begin{pmatrix}
2 \left( \lambda_1 + \lambda_3 \right) & 2 \left( \lambda_2 + \lambda_3 \right) & 2 \lambda_7 & \sqrt{2}\left( \lambda_4^\mathrm{R} + i \lambda_4^\mathrm{I} \right) \\
2 \left( \lambda_2 + \lambda_3 \right) & 2 \left( \lambda_1 + \lambda_3 \right) & 2 \lambda_7 & -\sqrt{2}\left( \lambda_4^\mathrm{R} + i \lambda_4^\mathrm{I} \right) \\
2 \lambda_7 & 2 \lambda_7 & 2 \lambda_8 & 0 \\
\sqrt{2}\left( \lambda_4^\mathrm{R} - i \lambda_4^\mathrm{I} \right) & \sqrt{2}\left(-\lambda_4^\mathrm{R} + i \lambda_4^\mathrm{I} \right)  & 0 & \lambda_5 + \lambda_6
\end{pmatrix},\\
S_2^0 &= \begin{pmatrix}
2 \left( \lambda_1 - \lambda_2 \right) & 2 \left( \lambda_4^\mathrm{R} + i \lambda_4^\mathrm{I} \right) \\
2 \left( \lambda_4^\mathrm{R} - i \lambda_4^\mathrm{I} \right) & \lambda_5 + \lambda_6
\end{pmatrix}.
\end{align}
\end{subequations}
The eigenvalues can be compared against those in Ref.~\cite{Das:2014fea}, which are denoted as $\{a_i^\pm,\, b_j\}$. The eigenvalues of $S_1^0$ are $\{a_1^\pm,\, a_2^\pm\}$ and those of $S_2^0$ are $a_1^\pm$.

For simplicity, we rotate the $S$-matrix constructed from the states $\Psi^0_3$ by
\begin{equation}
\mathcal{P} \mathcal{R}\,S_3^0\,\mathcal{R}^\dagger \mathcal{P}^\mathrm{T} = \diag \left( \begin{pmatrix}
\mathcal{N}_{11} & \mathcal{N}_{12} \\
\mathcal{N}_{12} & \mathcal{N}_{11}
\end{pmatrix},\,\mathcal{N}_{33} \right),
\end{equation} 
where the rotations are given by
\begin{subequations}
\begin{align}
\mathcal{R}={}&\mathcal{I}_5 \otimes U,\\
\mathcal{P}={}& \begin{pmatrix}
 1 & 0 & 0 & 0 & 0 & 0 & 0 & 0 & 0 & 0 \\
 0 & 0 & 1 & 0 & 0 & 0 & 0 & 0 & 0 & 0 \\
 0 & 0 & 0 & 1 & 0 & 0 & 0 & 0 & 0 & 0 \\
 0 & 0 & 0 & 0 & 1 & 0 & 0 & 0 & 0 & 0 \\
 0 & 0 & 0 & 0 & 0 & 0 & 1 & 0 & 0 & 0 \\
 0 & 0 & 0 & 0 & 0 & 0 & 0 & 1 & 0 & 0 \\
 0 & 1 & 0 & 0 & 0 & 0 & 0 & 0 & 0 & 0 \\
 0 & 0 & 0 & 0 & 0 & 1 & 0 & 0 & 0 & 0 \\
 0 & 0 & 0 & 0 & 0 & 0 & 0 & 0 & 1 & 0 \\
 0 & 0 & 0 & 0 & 0 & 0 & 0 & 0 & 0 & 1 
\end{pmatrix},
\end{align}
\end{subequations}
and the elements of the scattering matrix are then
\begin{subequations}\label{Eq:N11_N12_N33}
\begin{align}
\mathcal{N}_{11}=&{}\begin{pmatrix}
2\left( \lambda_1 + \lambda_2 + 2 \lambda_3 \right) & - 4i \lambda_4^\mathrm{I} &  4 \lambda_4^\mathrm{R}\\
4i \lambda_4^\mathrm{I} & \lambda_5 + \lambda_6 - 4 \lambda_7 & 0 \\
4 \lambda_4^\mathrm{R} & 0 & \lambda_5 + \lambda_6 + 4 \lambda_7
\end{pmatrix},\\
\mathcal{N}_{12}=&{}\begin{pmatrix}
4 \lambda_3 & - 2i \lambda_4^\mathrm{I} &  2 \lambda_4^\mathrm{R}\\
2i \lambda_4^\mathrm{I} & \lambda_6 - 2 \lambda_7 & 0 \\
2 \lambda_4^\mathrm{R} & 0 & \lambda_6 + 2 \lambda_7
\end{pmatrix},\\
\mathcal{N}_{33}=&{}\begin{pmatrix}
6 \lambda_1 - 2 \lambda_2 + 4 \lambda_3 & 4 \lambda_1 & \lambda_6 & 2\lambda_5 + \lambda_6\\
4 \lambda_1 & 6 \lambda_1 - 2 \lambda_2 + 4 \lambda_3 & -\lambda_6 & 2 \lambda_5 +\lambda_6\\
\lambda_6 & -\lambda_6 & 2 \lambda_8 & 0 \\
2\lambda_5 + \lambda_6 & 2\lambda_5 + \lambda_6 & 0 & 6 \lambda_8
\end{pmatrix}.
\end{align}
\end{subequations}
The eigenvalues of $\mathcal{N}_{33}$ are $\{a_3^\pm,\, a_5^\pm\}$. Due to the $\lambda_4^\mathrm{I}$ there is an additional mixing between the eigenvalues and those do not simplify to neither $a_i^\pm$ nor $b_i$. However, in the limit of $\lambda_4^\mathrm{I}=0$ we get $\{a_4^\pm,\, a_6^\pm,\, b_1,\, b_2\}$. These are the only eigenvalues which differ between the models with real and complex coefficients. As shall be discussed, see eq.~\eqref{Eq:New_S30_S40}, the block-diagonal structures can be simplified.

Finally, the $\Psi^0_4$ states can be rotated by
\begin{equation}
\mathcal{P} \mathcal{R}\,S_4^0\,\mathcal{R}^\dagger \mathcal{P}^\mathrm{T} = \diag \left(
\mathcal{N}_{44},\, \begin{pmatrix}
\mathcal{N}_{11} & \mathcal{N}_{12} \\
\mathcal{N}_{12} & \mathcal{N}_{11}
\end{pmatrix} \right),
\end{equation} 
where the rotation matrices are
\begin{subequations}
\begin{align}
\mathcal{R}={}&\mathcal{I}_4 \otimes U,\\
\mathcal{P}={}& \begin{pmatrix}
 1 & 0 & 0 & 0 & 0 & 0 & 0 & 0 \\
 0 & 0 & 0 & 0 & 1 & 0 & 0 & 0 \\
 0 & 1 & 0 & 0 & 0 & 0 & 0 & 0 \\
 0 & 0 & 1 & 0 & 0 & 0 & 0 & 0 \\
 0 & 0 & 0 & 1 & 0 & 0 & 0 & 0 \\
 0 & 0 & 0 & 0 & 0 & 1 & 0 & 0 \\
 0 & 0 & 0 & 0 & 0 & 0 & 1 & 0 \\
 0 & 0 & 0 & 0 & 0 & 0 & 0 & 1 
\end{pmatrix},
\end{align}
\end{subequations}
and the elements are
\begin{equation}
\mathcal{N}_{44} = \begin{pmatrix}
2 \lambda_1 - 6 \lambda_2 - 4 \lambda_3 & -4 \lambda_2 \\
-4 \lambda_2 & 2 \lambda_1 - 6 \lambda_2 - 4 \lambda_3
\end{pmatrix}.
\end{equation}
The eigenvalues of $\mathcal{N}_{44}$ are $\{b_3,\, b_4\}$. As will be discussed, see eq.~\eqref{Eq:New_S30_S40}, the block-diagonal structures can be simplified.

The singly-charged 2-body states can be grouped as follows,
\begin{subequations}
\begin{align}
\Psi^+_1 & = \Big\{ \left.| n_1 w^+_1 \right\rangle,\ \left.| n_2 w^+_2 \right\rangle,\ \left.| n_S w^+_S \right\rangle,\ \left.| n_S w^+_2 \right\rangle,\ \left.| n_2 w^+_S \right\rangle \Big\},\\
\Psi^+_2 &= \Big\{ \left.| n_1 w^+_2 \right\rangle,\ \left.| n_2 w^+_1 \right\rangle,\ \left.| n_1 w^+_S \right\rangle,\ \left.| n_S w^+_1 \right\rangle \Big\},\\
\Psi^+_3 &= \Big\{ \left.| n_1^\ast w^+_1 \right\rangle,\ \left.| n_2^\ast w^+_2 \right\rangle,\ \left.| n_S^\ast w^+_S \right\rangle,\ \left.| n_S^\ast w^+_2 \right\rangle,\ \left.| n_2^\ast w^+_S \right\rangle \Big\},\\
\Psi^+_4 &= \Big\{ \left.| n_1^\ast w^+_2 \right\rangle,\ \left.| n_2^\ast w^+_1 \right\rangle,\ \left.| n_1^\ast w^+_S \right\rangle,\ \left.| n_S^\ast w^+_1 \right\rangle \Big\}.
\end{align}
\end{subequations}

The first $S$-matrix can be rotated by
\begin{subequations}
\begin{align}
\mathcal{R}={}& \diag (U,\,1,\,U),\\
\mathcal{P}={}& \begin{pmatrix}
1 & 0 & 0 & 0 & 0\\
0 & 0 & 0 & 0 & 1\\
0 & 0 & 0 & 1 & 0\\
0 & 0 & 1 & 0 & 0\\
0 & 1 & 0 & 0 & 0 
\end{pmatrix},
\end{align}
\end{subequations}
which yields
\begin{equation}
\mathcal{P} \mathcal{R}\,S_1^+\,\mathcal{R}^\dagger \mathcal{P}^\mathrm{T} = \begin{pmatrix}
2 \left( \lambda_1 - \lambda_2 \right) & 2 \left( \lambda_4^\mathrm{R} + i \lambda_4^\mathrm{I} \right) & 0 & 0 & 0\\
2\left( \lambda_4^\mathrm{R} - i \lambda_4^\mathrm{I} \right) & \lambda_5 + \lambda_6 & 0 & 0 & 0\\
0 & 0 & \lambda_5 - \lambda_6 & 0 & 0\\
0 & 0 & 0 & 2\lambda_8 & 2 \sqrt{2} \lambda_7 \\
0 & 0 & 0 & 2\sqrt{2} \lambda_7 & 2\left( \lambda_1 + \lambda_2 + 2 \lambda_3 \right)
\end{pmatrix}.
\end{equation}
The eigenvalues are $\{a_1^\pm,\, a_2^\pm,\, b_6\}$.

Then, the $\Psi_2^+$ can be rotated by
\begin{subequations}
\begin{align}
\mathcal{R}={}& \mathcal{I}_2 \otimes U,\\
\mathcal{P}={}& \begin{pmatrix}
1 & 0 & 0 & 0 \\
0 & 0 & 1 & 0 \\
0 & 1 & 0 & 0 \\
0 & 0 & 0 & 1 \\
\end{pmatrix},
\end{align}
\end{subequations}
which results in 
\begin{equation}
\mathcal{P} \mathcal{R}\,S_2^+\,\mathcal{R}^\dagger \mathcal{P}^\mathrm{T} = \begin{pmatrix}
2\left( \lambda_1 + \lambda_2 - 2 \lambda_3 \right) & 0 & 0 & 0 \\
0 & \lambda_5 - \lambda_6 & 0 & 0 \\
0 & 0 & 2 \left( \lambda_1 - \lambda_2 \right) & 2 \left( \lambda_4^\mathrm{R} + i \lambda_4^\mathrm{I} \right)\\
0 & 0 & 2\left( \lambda_4^\mathrm{R} - i \lambda_4^\mathrm{I} \right) & \lambda_5 + \lambda_6
\end{pmatrix}.
\end{equation}
The eigenvalues are $\{a_1^\pm,\, b_5,\, b_6\}$.

For $\Psi_3^+$ we got 
\begin{subequations}
\begin{align}
\mathcal{R}={}& \diag (U,\,1,\,U),\\
\mathcal{P}={}& \begin{pmatrix}
0 & 0 & 1 & 0 & 0\\
0 & 1 & 0 & 0 & 0\\
1 & 0 & 0 & 0 & 0\\
0 & 0 & 0 & 1 & 0\\
0 & 0 & 0 & 0 & 1 
\end{pmatrix},
\end{align}
\end{subequations}
and
\begin{equation}
\mathcal{P} \mathcal{R}\,S_3^+\,\mathcal{R}^\dagger \mathcal{P}^\mathrm{T} = \diag \left( \begin{pmatrix}
2 \lambda_8 & \sqrt{2} \lambda_6 \\
\sqrt{2} \lambda_6 & 2\left( \lambda_1 - \lambda_2 + 2 \lambda_3 \right)
\end{pmatrix},\, \mathcal{N}_{55} \right),
\end{equation}
where
\begin{equation}
\mathcal{N}_{55} = \begin{pmatrix}
2 \left( \lambda_1 + \lambda_2 \right) & -2i \lambda_4^\mathrm{I} & 2\lambda_4^\mathrm{R}\\
2i\lambda_4^\mathrm{I} & \lambda_5 - 2 \lambda_7 & 0 \\
2\lambda_4^\mathrm{R} & 0 & \lambda_5 + 2 \lambda_7
\end{pmatrix}.
\end{equation}
The eigenvalues of the first block are $a_3^\pm$, while those of $\mathcal{N}_{55}$ in the limit of $\lambda_4^\mathrm{I}=0$ coincide with $\{a_4^\pm,\, b_2 \}$. The three eigenvalues of $\mathcal{N}_{55}$ should correspond to the eigenvalues coming from the first diagonal block of $S_3^0$ given by eq.~\eqref{Eq:N11_N12_N33}. This observations lets us conclude that the $6\times6$ matrix~\eqref{Eq:N11_N12_N33} can be split into two block-diagonal structures of dimension three; one corresponding to $\mathcal{N}_{55}$. By observing the structure of $\mathcal{N}_{55}$ and comparing the eigenvalues we can guess that the other block should correspond to
\begin{equation}
\mathcal{N}_{66} = \begin{pmatrix}
2 \left( \lambda_1 + \lambda_2 + 4 \lambda_3 \right) & -6i \lambda_4^\mathrm{I} & 6\lambda_4^\mathrm{R}\\
6i\lambda_4^\mathrm{I} & \lambda_5 + 2 \lambda_6 - 6 \lambda_7 & 0 \\
6\lambda_4^\mathrm{R} & 0 & \lambda_5 + 2 \lambda_6 + 6 \lambda_7
\end{pmatrix}.
\end{equation}
In spite of not knowing how the rotation matrix looks like, the eigenvalues of the $S$-matrix are what we care about when analysing the unitarity conditions. We conclude that after some rotation $\mathcal{R}_i$ the $S_i^0$-matrices are given by
\begin{subequations}\label{Eq:New_S30_S40}
\begin{align}
\mathcal{R}_3 \,S_3^0 \,\mathcal{R}_3^\dagger ={}& \diag \left( \mathcal{N}_{33},\, \mathcal{N}_{55},\, \mathcal{N}_{66} \right),\\
\mathcal{R}_4 \,S_4^0 \,\mathcal{R}_4^\dagger ={}& \diag \left( \mathcal{N}_{44},\, \mathcal{N}_{55},\, \mathcal{N}_{66} \right).
\end{align}
\end{subequations}

Finally, the last piece of the singly-charged $S$-matrix can be rotated by
\begin{equation}
\mathcal{R}= \mathcal{I}_2 \otimes U,
\end{equation}
so that 
\begin{equation}
\mathcal{R}\,S_4^+\,\mathcal{R}^\dagger = \diag \left( 2\left( \lambda_1 - \lambda_2 - 2 \lambda_3 \right),\, \mathcal{N}_{55}^\ast \right).
\end{equation}
The first entry coincides with $b_4$ while eigenvalues from $\mathcal{N}_{55}$ in the limit of $\lambda_4^\mathrm{I}=0$ coincide with $\{a_4^\pm,\, b_2 \}$.

In general, the doubly-charged $S$-matrices can be omitted as they yield redundant eigenvalues. For completeness, we list these states:
\begin{subequations}
\begin{align}
\Psi^{++}_1 &= \Big\{ \left.| w^+_1 w^+_1 \right\rangle,\, \left.| w^+_2 w^+_2 \right\rangle,\, \left.| w^+_S w^+_S \right\rangle,\ \left.| w^+_2 w^+_S \right\rangle \Big\},\\
\Psi^{++}_2 &= \Big\{ \left.| w^+_1 w^+_2 \right\rangle,\, \left.| w^+_1 w^+_S \right\rangle \Big\}.
\end{align}
\end{subequations}
The scattering matrices are
\begin{subequations}
\begin{align}
S_1^{++} &= S_1^0,\\
S_2^{++} &= S_2^0.
\end{align}
\end{subequations}

\bibliographystyle{JHEP}

\bibliography{ref}

\end{document}